\documentclass[reprint,
groupedaddress,
amsmath,
amssymb,
aps,
prb,
floatfix,
]{revtex4-2}

\usepackage[justification=Justified, format=plain, font=small]{caption}
\usepackage[justification=Justified, format=plain, font=small]{subcaption}

\usepackage{braket}
\usepackage{ulem}
\usepackage{appendix}

\usepackage[pdftex]{graphicx}
\usepackage{dcolumn}
\usepackage{bm}
\usepackage[pdftex, colorlinks=true, linkcolor=blue, urlcolor=blue,citecolor=blue]{hyperref}

\usepackage{color}
\usepackage{here}
\usepackage{txfontsb}
\usepackage{mathcomp}
\usepackage{dsfont}
\usepackage{physics}
\usepackage{amsmath,amssymb,mathtools,bm}
\usepackage{enumitem}
\usepackage{bbm}
\usepackage{mathrsfs}

\definecolor{mygray}{gray}{0.7}

\DeclareMathOperator{\TrOp}{Tr}

\hypersetup{colorlinks=true,linkcolor=blue,citecolor=blue,urlcolor=blue}

\begin{document}

\title{Irreducible Geometry of Higher-Order Correlator Families}
  \author{Kaito Kobayashi}
  \email{kaito-kobayashi@aion.t.u-tokyo.ac.jp}
  \affiliation{Department of Applied Physics, the University of Tokyo, Tokyo 113-8656, Japan}
\date{\today}

\begin{abstract}
    Programmable quantum simulators are beginning to access correlators of increasing complexity, ranging from four-point out-of-time-ordered correlators to even higher-order many-body correlators. 
    The theoretical framework for interpreting such data, however, remains comparatively underdeveloped. 
    Although a variety of higher-order correlators can be constructed straightforwardly by changing the ordering, timing, or choice of operators, their physical meaning is often difficult to infer.
    A further complication is that different correlators are generally not independent: some may be mutually redundant, while others may encode genuinely distinct many-body information.
    These features make it necessary to analyze correlators not as isolated quantities, but as a structured family.
    In this work, we develop a geometric framework for the collective analysis of higher-order correlator families. 
    By representing correlators as inner products between operator words, we recast each family as a geometry in operator space. 
    The key idea is to introduce conditioning subspaces that separate this geometry into reducible information, already explained by a chosen resolved sector, and irreducible information, encoded in the residual correlator geometry.
    Focusing on the latter component, we define irreducible volume profiles that quantify how broadly the unexplained part of a correlator family spreads over independent geometric directions. 
    This perspective leads to several complementary forms of conditioning. 
    Canonical conditioning optimally explains a correlator family, revealing that free-fermion, interacting integrable, and chaotic spin-chain dynamics generate qualitatively different correlator geometries. 
    Targeted conditioning instead fixes the resolved sector to isolate a chosen physical feature, allowing us to identify the spatial confinement of correlator geometry under many-body localization, characterize the organization of measurement-inaccessible correlator components, and reveal state-dependent structure in correlator geometry from a spectral perspective.
    Finally, Krylov and cross conditioning extend the framework from a single correlator family to comparisons among correlator geometries: the former tracks how the geometry evolves in time, while the latter identifies irreducible structures that are shared or reshaped across different dynamics.
    Our framework reveals irreducible structures hidden at the level of individual correlator values and establishes correlator geometry as a higher-level description of quantum many-body dynamics.
\end{abstract}

\maketitle

\section{Introduction}

Correlators are the primary language for probing quantum many-body physics. 
In their simplest form, equal-time correlators \(\langle O_i O_j\rangle\) quantify spatial correlations and often serve as order parameters for equilibrium phases~\cite{Sachdev:Cambridge:2011}. 
Allowing operators to evolve in the Heisenberg picture gives dynamical correlators \(\langle O_i (t) O_j\rangle\), which reveal the dynamics of quantum matter through response functions and excitation spectra~\cite{VanHove:PR:1954,Kubo:JPSJ:1957}. 
The Heisenberg evolution also provides a natural framework for operator spreading, in which an initially local operator develops support over an increasingly extended set of degrees of freedom~\cite{Lieb:1972,Zanardi:PRA:2001,Prosen:PRA:2007,Pizorn:PRB:2009,Zhou:PRB:2017,Keyserlingk:PRX:2018,Nahum:PRX:2018,Lin:PRB:2018,Kobayashi:SciPostPhys:2025}. 
Out-of-time-ordered correlators (OTOCs) \(\langle O_i(t)O_jO_i(t)O_j\rangle\) provide a direct probe of this spreading and have become standard diagnostics of quantum chaos, scrambling, and quantum gravity~\cite{Larkin:1969,Shenker:JHEP:2014,Roberts:JHEP:2015,Maldacena:JHEP:2016,Hosur:JHEP:2016,Hashimoto:JHEP:2017,Cotler:AnnPhys:2018,Swingle:NatPhys:2018,Khemani:PRX:2018,Rakovszky:PRX:2018,Xu:PRX:2019}. 
On the experimental side, correlators have become increasingly accessible as measurable observables. 
Beyond equal-time and dynamical correlators that have long been measured in condensed matter physics~\cite{Shirane:RevModPhys:1974,Dai:RevModPhys:2015,Sebastian:RevModPhys:2019}, OTOCs have now been accessed across a range of quantum platforms~\cite{Swingle:PRA:2016,Yao:arXiv:2016,Jun:PRX:2017,Garttner:NatPhys:2017,Wei:PRL:2018,Landsman:Nature:2019,Vermersch:PRX:2019,Joshi:PRL:2020,Mi:Science:2021,Pegahan:PRL:2021,Green:PRL:2022,Braumuller:NatPhys:2022,Zhang:arXiv:2025}. 
A further step was reported by Google Quantum AI, which measured higher-order OTOCs \(\langle (O_i(t)O_j)^{2n}\rangle\) on a large-scale superconducting processor~\cite{Google:Nature:2025}. 
These developments mark a new stage in quantum many-body dynamics, in which programmable quantum devices can access correlators of unprecedented complexity.

\begin{figure}[b!]
    \centering
    \includegraphics[width=\hsize]{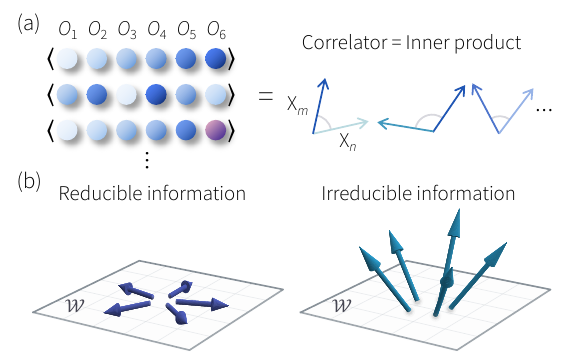}
    \caption{
        (a) Schematic illustration of a family of correlators. 
        A variety of higher-order correlators can be generated either by permuting the operators or by replacing some of them with others.
        Geometrically, these correlators can be viewed as inner products of operator words. 
        (b) Correlator decomposition into the reducible part (the inner product of the projected components) and the irreducible part (the inner product of the residual components) relative to the conditioning subspace \(\mathcal{W}\).
    }\label{fig1}
\end{figure}

At the same time, this progress exposes a growing gap between increasingly complex correlator measurements and the theoretical framework needed to organize and interpret them.
Although higher-order correlators have been analyzed in several relatively structured settings~\cite{Roberts:JHEP:2017,Haehl:SciPostPhys:2019,Fujii:arXiv:2025,Fritzsch:arXiv:2025}, a general framework for characterizing more intricate and less constrained families of correlators remains largely unexplored.
Consider, for instance, a setting in which arbitrary six-point correlators can be measured.
Even for a fixed set of operators, many distinct correlators can be generated by permuting their order; allowing some operators to be replaced by others generates many more, as illustrated in Fig.~\ref{fig1}(a).
Notably, the resulting collection of correlators exhibits varying degrees of dependence and redundancy among its members.
For example, one may compare \(\langle O_1O_2O_3O_4O_5O_6\rangle\), \(\langle O_1O_2O_3O_4O_6O_5\rangle\), and \(\langle O_1O_2O_3O_4O_5O_6'\rangle\). 
The first and second are redundant when \(O_5\) and \(O_6\) commute, whereas the first and third are related but not equivalent when \(O_6'\) is a small deformation of \(O_6\). 
More generally, different correlators can encode genuinely distinct many-body information.
Thus, selecting one correlator as canonical is physically unjustified in general, while simply tabulating hundreds of raw correlator values can fill pages without revealing the organizing principles behind the data. 
A systematic framework is therefore needed to analyze higher-order correlators collectively, rather than as isolated observables, and to identify the physics encoded in their mutual organization.

In this work, we propose a geometric framework for the collective analysis of higher-order correlator families. 
The construction begins with the observation that any correlator can be represented as an inner product \(\langle X_m, X_n\rangle\) between two operator words in operator space [Fig.~\ref{fig1}(a)]. 
Here, an operator word may be a single operator \(X_m=O_i\) or a composite product \(X_m=O_iO_j\cdots\). 
Given a collection of operator words \(\Omega =\{X_m\}\), we define the correlator family of interest as the set of all pairwise inner products among the elements of \(\Omega\). 
Analyzing this correlator family therefore amounts to studying the geometry spanned by these operator words in operator space.

Within this geometric perspective, relations among correlators are characterized by decomposing the correlator geometry relative to a conditioning subspace \(\mathcal{W}\).
Here, \(\mathcal{W}\) defines the \textit{resolved} sector of operator space, whose contribution to the correlator family is treated as already accounted for.
Redundancy and independence are then assessed relative to this resolved sector.
More concretely, once \(\mathcal{W}\) is fixed, each operator word is decomposed into its projection onto \(\mathcal{W}\) and the corresponding residual, as illustrated in Fig.~\ref{fig1}(b).
Inner products among the projected components give the part of the correlator family captured by the resolved sector \(\mathcal{W}\).
We therefore focus on the residuals, since their mutual inner products define the irreducible component of the correlator family left unexplained by \(\mathcal{W}\).
If these residuals span only a few nearly redundant directions, the unexplained part carries largely overlapping information; if they span many geometrically independent directions, it encodes a broader set of independent irreducible information.
We quantify this structure by the volumes spanned by the residuals at each dimensional level, which yields a dimension-resolved profile of irreducible correlator geometry relative to \(\mathcal{W}\).

We apply this framework to three complementary choices of conditioning subspace, each designed to probe a different aspect of the irreducible correlator geometry.
First, we define canonical conditioning, in which \(\mathcal{W}\) is chosen to optimally explain the given correlator family by minimizing the irreducible component outside \(\mathcal{W}\). 
Applying this construction to quantum spin chains, we find that free-fermion, interacting integrable, and chaotic dynamics generate qualitatively different correlator geometries. 
Second, we introduce targeted conditioning, in which the conditioning subspace is chosen to probe a specific physical characteristic. 
We use this construction to demonstrate the spatial confinement of correlator geometry induced by many-body localization (MBL)~\cite{Rahul:AnnRev:2015}, analyze the structure of measurement-inaccessible correlator components, and reveal state dependence encoded in the geometry from a spectral perspective.
Third, we introduce Krylov conditioning and cross conditioning to compare correlator geometries both across time and across different dynamics.
The former tracks how correlator geometry evolves in time, while the latter identifies which irreducible structures are shared, enhanced, or suppressed when the underlying dynamics changes.

In this way, our framework turns the proliferation of higher-order correlators from a challenge of interpretation into a diagnostic of quantum many-body dynamics.
Higher-order correlators are no longer viewed as isolated observables; instead, they are organized into a correlator geometry whose irreducible structure reveals many-body physics inaccessible from individual correlator values alone.

The remainder of this paper is organized as follows.
In Sec.~\ref{sec:2}, we develop the formalism of correlator geometry by introducing operator Gram matrices and residual geometries.
In Sec.~\ref{sec:3}, we define canonical conditioning as an intrinsic choice of conditioning sector. We first establish analytic benchmarks using Pauli-string families and Haar-randomized operator words, and then apply the construction to free-fermion, interacting integrable, and chaotic spin-chain dynamics.
In Sec.~\ref{sec:4}, we present targeted conditioning and apply it to spatial confinement in MBL dynamics, measurement accessibility, and state dependence in correlator geometry.
In Sec.~\ref{sec:5}, we introduce Krylov conditioning and cross conditioning to compare how irreducible structures evolve in time and change across dynamics.
Section~\ref{sec:6} is devoted to conclusions.

\section{Correlator geometry}
\label{sec:2}

We first develop the basic formalism of correlator geometry.
In Sec.~\ref{sec:2:subsec:A}, we introduce the operator Gram matrix, which recasts a correlator family as the pairwise geometry of operator words.
In Sec.~\ref{sec:2:subsec:B}, we formulate a conditioning procedure based on a subspace \(\mathcal{W}\), thereby decomposing this geometry into a component resolved by \(\mathcal{W}\) and a residual component left unexplained by \(\mathcal{W}\).

\subsection{Operator Gram matrix}
\label{sec:2:subsec:A}

Let \(\mathcal{H}\) be a Hilbert space of dimension \(d\), and consider the operator space \(\mathcal{B}(\mathcal{H})\) equipped with the Hilbert-Schmidt inner product
\begin{equation}
    \langle X,Y\rangle_{\mathrm{HS}}=\frac{1}{d}\TrOp(X^{\dagger}Y).
    \label{eq:HS_inner}
\end{equation}
The associated Hilbert-Schmidt norm is \(\|X\|_{\mathrm{HS}}^2=\langle X, X\rangle_{\mathrm{HS}}\). 
We consider operator words \(X_m\in \mathcal{B}(\mathcal{H})\), which may be single Heisenberg operators \(O_j(t_j)\) or products of multiple operators \(O_i(t_i)O_j(t_j)\cdots\). 
For notational simplicity, we normalize each word as \(\hat{X}_m=X_m/\|X_m\|_{\mathrm{HS}}\) and omit the hat in what follows.

For a finite collection of normalized operator words
\begin{equation}
    \Omega = \{X_m\}_{m=1}^N,
\end{equation}
we define the associated correlator family as the set of all pairwise overlaps among the elements of \(\Omega\). 
We collect these overlaps into a matrix, called the Gram matrix,
\begin{equation}
    G_{mn}(\Omega)=\langle X_m,X_n\rangle_{\mathrm{HS}}
    =\frac{1}{d}\TrOp(X_m^{\dagger}X_n).
\end{equation}
By choosing the operator words \(X_m\) appropriately, the entries of \(G(\Omega)\) can represent ordinary time-ordered correlators, OTOCs, or more general higher-order correlators. 
Examples of Gram matrices constructed from higher-order OTOC families are given in Appendix~\ref{Appendix:OTOCs}.

The Gram matrix has a direct geometric meaning.
Let \(\operatorname{Vol}(X_1,\dots,X_N)\) denote the \(N\)-dimensional volume spanned by the selected operator words. 
Then the determinant of the Gram matrix is equal to the square of this volume:
\begin{equation}
    \det G(\Omega)=\operatorname{Vol}^2(X_{1},\dots,X_{N}) \geq 0.
    \label{eq:gram_det_volume}
\end{equation}
A small determinant indicates that the selected operator words are nearly linearly dependent, while a large determinant indicates that they span many independent directions. 
Note that the same volume interpretation extends to any subfamily.
For a subset of indices \(\{i_1,\dots,i_q\}\), we define \(\Omega_q=\{X_{i_1},\dots,X_{i_q}\}\subseteq \Omega\). 
Then \(G(\Omega_q)\) is precisely the principal submatrix of \(G(\Omega)\) obtained by retaining the rows and columns indexed by \(\{i_1,\dots,i_q\}\).  
The corresponding principal minor is given by \(\det G(\Omega_q)\), which is equal to the squared \(q\)-dimensional volume spanned by the selected operator words.

It is useful to show a simple example here. 
Consider the set of three operator words \(\Omega = \{X_m\}_{m=1}^3\), defined by
\begin{equation}
    X_1=U_1U_2U_3,\quad X_2=U_2U_1U_3,\quad X_3=U_2U_3U_1,
\end{equation}
where each \(U_i\) is unitary.
The corresponding Gram matrix is
\begin{equation}
    G(\Omega)=\begin{pmatrix}
        1 & f_4 & h_6\\
        f_4^{*} & 1 & g_4\\
        h_6^{*} & g_4^{*} & 1
    \end{pmatrix},
\end{equation}
where
\begin{equation}
    \begin{aligned}
        f_4 &=\langle X_1, X_2\rangle = \frac{1}{d}\TrOp(U_2^\dagger U_1^\dagger U_2 U_1), \\
        g_4 &= \langle X_2, X_3\rangle =\frac{1}{d}\TrOp(U_3^\dagger U_1^\dagger U_3 U_1),\\
        h_6 &= \langle X_1, X_3\rangle =\frac{1}{d}\TrOp(U_3^\dagger U_2^\dagger U_1^\dagger U_2 U_3 U_1). 
    \end{aligned}\label{eq:f4_g4_h6}
\end{equation}
By unitarity and cyclicity of the trace, \(f_4\) and \(g_4\) reduce to four-point correlators, whereas \(h_6\) is a six-point correlator involving all three unitaries.

The determinant can be evaluated explicitly for this small set of operator words, yielding a simple nontrivial constraint. 
Specifically, the positivity condition \(\det G(\Omega)\ge 0\) implies
\begin{equation}
    |h_6-f_4g_4|
    \le
    \sqrt{(1-|f_4|^2)(1-|g_4|^2)}.
    \label{eq:3word_disk}
\end{equation}
Thus, once \(f_4\) and \(g_4\) are fixed, the six-point correlator \(h_6\) is constrained to lie inside a closed disk in the complex plane. 
The center and radius of this disk are determined entirely by the two four-point correlators.

This example already captures the basic motivation of our work.
First, consider a two-qubit system with
\begin{equation}
    U_1=\sigma^z\otimes \mathbb{I},\quad
    U_2=\mathbb{I}\otimes \sigma^z,\quad
    U_3=R_x(a) \otimes \mathbb{I},
\end{equation}
where \(\sigma^\alpha\) are the Pauli operators for \(\alpha=x,y,z\), \(\mathbb{I}\) is the identity, and \(R_{\alpha}(\theta)=e^{-i\theta \sigma^\alpha/2}\).
A direct calculation gives
\begin{equation}
    f_4=1,\quad
    g_4=\cos(a),\quad
    h_6=\cos(a).
\end{equation}
Since \(f_4=1\), the disk radius in Eq.~\eqref{eq:3word_disk} vanishes. 
Therefore, \(h_6\) is not an independent quantity: once \(f_4\) and \(g_4\) are specified, \(h_6\) is fixed.

Now consider a second realization, again on two qubits:
\begin{equation}
    U_1=\sigma^z\otimes\mathbb{I},
    \quad
    U_2=R_x(b)\otimes\mathbb{I},
    \quad
    U_3=R_y(c)\otimes\mathbb{I} .
\end{equation}
In this case,
\begin{equation}
    f_4=\cos(b),\quad
    g_4=\cos(c),\quad
    h_6=\cos(b)\cos(c).
\end{equation}
If the parameters are chosen such that \(\cos(a)=\cos(b)\cos(c)\), then the raw value of \(h_6\) is the same in the two realizations.
However, in the second realization, the disk radius is generically nonzero, so the same value of \(h_6\) lies within a finite allowed region rather than being uniquely fixed.
This distinction highlights the need for a collective treatment of correlators, beyond an analysis based on individual values alone.

Note that such a direct interpretation is possible only when \(\Omega\) is small.
Once the correlator family becomes large, the determinant of the Gram matrix combines many correlators at once, making it difficult to identify which aspects of the correlator family are already accounted for and which remain genuinely independent.
To make this distinction well defined, we introduce a conditioning subspace \(\mathcal{W}\), which specifies the sector of operator space treated as \textit{resolved}.

\subsection{Reducible and irreducible information}
\label{sec:2:subsec:B}

We now define a conditioning procedure relative to a chosen subspace \(\mathcal{W}\).
Let \(\mathbb{P}_{\mathcal{W}}:\mathcal{B}(\mathcal{H})\to\mathcal{B}(\mathcal{H})\) be the orthogonal projector onto \(\mathcal{W}\).
For each operator word \(X_m\), we write
\begin{equation}
    X_m=\mathbb{P}_{\mathcal{W}}X_m+R_m^{(\mathcal{W})},
    \quad
    R_m^{(\mathcal{W})}\perp \mathcal{W}.
    \label{eq:qarm_residual_decomposition}
\end{equation}
The first term is the component projected onto \(\mathcal{W}\), while \(R_m^{(\mathcal{W})}\) is the residual component orthogonal to \(\mathcal{W}\).
By this orthogonality, the cross terms vanish, and the Gram matrix splits as
\begin{equation}
    G(\Omega)=\hat{G}^{(\mathcal{W})}(\Omega)+Q^{(\mathcal{W})}(\Omega),
    \label{eq:qarm_gram_split}
\end{equation}
where
\begin{align}
    \hat{G}_{mn}^{(\mathcal{W})}(\Omega)&=\langle \mathbb{P}_{\mathcal{W}}X_m,\mathbb{P}_{\mathcal{W}}X_n\rangle,\\
    Q_{mn}^{(\mathcal{W})}(\Omega)&=\langle R_m^{(\mathcal{W})},R_n^{(\mathcal{W})}\rangle.
    \label{eq:qarm_def_C_and_tildeG}
\end{align}
Note that both \(\hat{G}^{(\mathcal{W})}\) and \(Q^{(\mathcal{W})}\) are also Gram matrices: the former is formed from the projected operator words \(\{\mathbb{P}_{\mathcal{W}}X_m\}\), while the latter is obtained from the residual operator words  \(\{R_m^{(\mathcal{W})}\}\).
Therefore, the volume interpretation of determinants applies to both matrices.

The conditioning subspace \(\mathcal{W}\) specifies the sector of operator space that is treated as already \textit{resolved}. 
Operationally, this means that the projected component \(\mathbb{P}_{\mathcal{W}} X_m\) is assumed to be available for each \(X_m\), either by applying the projector \(\mathbb{P}_{\mathcal{W}}\) directly or by using the overlaps \(\langle Y_\alpha,X_m\rangle\) with an orthonormal basis \(\{Y_\alpha\}\) of \(\mathcal{W}\).
At this stage, we keep \(\mathcal{W}\) abstract.
Its physical interpretation depends on how \(\mathcal{W}\) is chosen: it may represent a purely mathematical subspace, a resource constraint, a coarse-grained description, a variational ansatz, a code subspace, a symmetry sector, a low-energy manifold, a hydrodynamic description, an effective theory, or an experimental constraint.
Once \(\mathcal{W}\) is fixed, \(\hat{G}^{(\mathcal{W})}\) describes the geometry of the reducible part, while \(Q^{(\mathcal{W})}\) captures the geometry of the irreducible components relative to the chosen conditioning sector.

To quantify this irreducible geometry, we first consider the total residual norm
\begin{equation}
    I_1(\Omega,\mathcal{W})=\sum_{m=1}^N\|R_m^{(\mathcal{W})}\|^2.
\end{equation}
Since the operator words \(X_m\) are normalized, \(0\le I_1(\Omega,\mathcal{W})\le N\). 
The lower bound is reached when every operator word lies inside \(\mathcal{W}\), while the upper bound is reached when every word is orthogonal to \(\mathcal{W}\). 
Thus, \(I_1\) quantifies the total weight of the correlator family remaining outside the resolved sector. 
By itself, however, \(I_1\) does not describe how that unresolved weight is organized.
For instance, the residuals may have large norms while being nearly collinear. 
In that case, \(I_1\) is large, yet the residual structure is highly redundant: adding a single direction to \(\mathcal{W}\) would account for most of it. 
Therefore, we need diagnostics that probe not only the total residual weight, but also the dimensional structure spanned by the residuals.

The first geometric diagnostic is the pairwise residual geometry.
For two operator words \(X_m\) and \(X_n\), let \(\Omega_{q=2}^{(m,n)}=\{X_m,X_n\}\). 
The determinant of the residual Gram matrix \(\det Q^{(\mathcal{W})}(\Omega_{q=2}^{(m,n)})\) is the squared area spanned by the two residuals \(R_m^{(\mathcal{W})}\) and \(R_n^{(\mathcal{W})}\). 
This determinant becomes small either when one of the residual norms is small, or when the two residuals are nearly aligned and hence geometrically redundant.
Conversely, a large determinant indicates that the pair contributes two linearly independent residual directions relative to \(\mathcal{W}\).
Summing over all pairs gives the second irreducible volume 
\begin{equation}
    I_2(\Omega,\mathcal{W})=\sum_{m<n}\det Q^{(\mathcal W)}(\Omega_{q=2}^{(m,n)}).
\end{equation}
A large value of \(I_2\) indicates that linearly independent residual pairs are widespread throughout the correlator family, rather than confined to a small number of exceptional pairs.

The construction extends naturally to general \(q\). 
For a subfamily \(\Omega_q=\{X_{i_1},\dots,X_{i_q}\}\subseteq\Omega\), the determinant of the residual Gram matrix \(Q^{(\mathcal{W})}(\Omega_q)\) is the squared \(q\)-dimensional volume spanned by the corresponding residuals. 
We define the irreducible volume by summing these determinants over all subfamilies of size \(q\):
\begin{equation}
    I_q(\Omega,\mathcal{W})
    =\sum_{\substack{\Omega_q\subseteq\Omega\\ |\Omega_q|=q}}\det Q^{(\mathcal{W})}(\Omega_q).
    \label{eq:def_residual_Iq}
\end{equation}
Thus, \(I_q(\Omega,\mathcal{W})\) quantifies how broadly \(q\)-dimensional irreducible structure is present across the residual family.

As discussed in Sec.~\ref{sec:2:subsec:A}, \(Q^{(\mathcal{W})}(\Omega_q)\) is a \(q\times q\) principal submatrix of \(Q^{(\mathcal{W})}(\Omega)\), and \(\det Q^{(\mathcal{W})}(\Omega_q)\) is the corresponding principal minor. 
Therefore, \(I_q(\Omega,\mathcal{W})\) is the sum of all \(q\times q\) principal minors of \(Q^{(\mathcal{W})}(\Omega)\). 
A standard result in linear algebra states that this sum equals the \(q\)-th elementary symmetric polynomial of the eigenvalues. 
In particular, denoting the eigenvalues of \(Q^{(\mathcal{W})}(\Omega)\) by \(\lambda_1,\dots,\lambda_N\ge0\), including possible zero eigenvalues, we obtain
\begin{equation}
    I_q(\Omega,\mathcal{W})=\sum_{1\le i_1<\cdots<i_q\le N}\lambda_{i_1}\cdots\lambda_{i_q}.
    \label{eq:Dq_elementary_symmetric}
\end{equation}
Since all eigenvalues are nonnegative, \(I_q(\Omega,\mathcal{W})>0\) if and only if \(\operatorname{rank}Q^{(\mathcal{W})}(\Omega)\ge q\).

To compare residual geometries across different settings, we normalize the sequence of volumes and define the irreducible volume profile
\begin{equation}
    \pi_q(\Omega,\mathcal{W})=
    \frac{I_q(\Omega,\mathcal{W})}{\sum_{q=0}^{N}I_q(\Omega,\mathcal{W})},
    \label{eq:total_residual_hierarchy_weight}
\end{equation}
where we set \(I_0(\Omega,\mathcal{W})=1\). 
This profile gives a dimension-resolved description of the residual correlator geometry: concentration near small \(q\) indicates that the residual geometry is organized by only a few independent directions, whereas broad support extending to larger \(q\) indicates that it spans a higher-dimensional set of directions.
We also use the mean and variance
\begin{equation}
    \bar{q}=\sum_{q=0}^{N} q\,\pi_q(\Omega,\mathcal{W}),
    \quad
    (\Delta q)^2=\sum_{q=0}^{N}(q-\bar{q})^2\pi_q(\Omega,\mathcal{W}),
    \label{eq:irreducible_order_variance}
\end{equation}
as a compact summary of the normalized profile.

\section{Canonical conditioning geometry}
\label{sec:3}

In this section, we introduce canonical conditioning, an intrinsic choice of conditioning sector determined by the operator word family itself.
Specifically, given a fixed family \(\Omega\), we choose the \(r\)-dimensional subspace \(\mathcal{W}\) that captures as much of \(\Omega\) as possible, or equivalently minimizes the total residual weight \(I_1(\Omega,\mathcal{W})\). 
We refer to this optimal construction as \(r\)-dimensional canonical conditioning.
The formal definition is given in Sec.~\ref{sec:3:subsec:A}. 
In Secs.~\ref{sec:3:subsec:B} and \ref{sec:3:subsec:C}, we provide two analytical benchmarks for Pauli-string families and Haar-randomized operator words, respectively.
In Sec.~\ref{sec:3:subsec:D}, we apply the construction numerically to free-fermion, interacting integrable, and chaotic spin-chain dynamics.

\subsection{Canonical conditioning}
\label{sec:3:subsec:A}

Let \(\Omega=\{X_m\}_{m=1}^{N}\) be a fixed normalized operator word family.
For a conditioning subspace \(\mathcal{W}\), the total residual weight is
\begin{equation}
    I_1(\Omega,\mathcal{W})
    =\sum_{m=1}^{N}\|R_{m}^{(\mathcal{W})}\|^2
    =\sum_{m=1}^{N}\|(\mathbb{I}-\mathbb{P}_{\mathcal{W}})X_{m}\|^2 .
    \label{eq:I1_general_principal}
\end{equation}
The \(r\)-dimensional canonical conditioning sector \(\mathcal{W}_r^\star\) is defined as the \(r\)-dimensional subspace that minimizes \(I_1\), or equivalently maximizes the total projected weight \(\sum_{m=1}^{N}\|\mathbb{P}_{\mathcal{W}}X_m\|^2\).
We now show that \(\mathcal{W}_r^\star\) is obtained from the leading \(r\) principal modes of the Gram matrix.

To begin with, we diagonalize the Gram matrix as
\begin{equation}
    G(\Omega)=V\operatorname{diag}(\lambda_1,\dots,\lambda_N)\,V^\dagger,
    \quad
    \lambda_1\ge\cdots\ge\lambda_N\ge0.
    \label{eq:principal_gram_decomp_simple}
\end{equation}
For every nonzero eigenvalue \(\lambda_\alpha>0\), we define the corresponding principal mode
\begin{equation}
    Y_\alpha=\frac{1}{\sqrt{\lambda_\alpha}}\sum_{m=1}^{N}V_{m\alpha}X_m.
    \label{eq:def_principal_modes_simple}
\end{equation}
These modes are orthonormal:
\begin{equation}
    \langle Y_\alpha,Y_\beta\rangle
    =\frac{(V^\dagger G(\Omega)V)_{\alpha\beta}}{\sqrt{\lambda_\alpha\lambda_\beta}}=\delta_{\alpha\beta},
    \label{eq:principal_orthonormal_simple}
\end{equation}
for all \(\alpha,\beta\) with nonzero eigenvalues.
Conversely, each operator word can be expanded as
\begin{equation}
    X_m=\sum_{\alpha}\sqrt{\lambda_\alpha}V_{m\alpha}^*Y_\alpha ,\label{eq:word_expansion_principal_simple}
\end{equation}
where the sum is over nonzero eigenvalues unless otherwise stated.

For each principal mode \(Y_\alpha\), the corresponding eigenvalue \(\lambda_\alpha\) is the total projected weight onto that mode:
\begin{equation}
    \sum_{m=1}^{N}
    \|\mathbb{P}_{Y_\alpha}X_m\|^2=
    \sum_{m=1}^{N}
    |\langle Y_\alpha,X_m\rangle|^2=\lambda_\alpha .
\end{equation}
Therefore, for a general \(r\)-dimensional subspace \(\mathcal{W}\), the captured weight can be written as
\begin{equation}
    \sum_{m=1}^{N}\|\mathbb{P}_{\mathcal{W}}X_m\|^2=\sum_\alpha \lambda_\alpha p_\alpha ,
\end{equation}
where 
\begin{equation}
  p_\alpha=\|\mathbb{P}_{\mathcal{W}}Y_\alpha\|^2,
  \quad
  0\le p_\alpha\le 1,
  \quad
  \sum_\alpha p_\alpha\le r.
\end{equation}
Since the eigenvalues are ordered, this expression is maximized by taking \(p_1=\cdots=p_r=1\) and \(p_\alpha=0\) for \(\alpha>r\), up to the usual freedom inside degenerate eigenspaces. 
Therefore, the optimal \(r\)-dimensional canonical conditioning sector is
\begin{equation}
    \mathcal{W}_r^\star=\operatorname{span}\{Y_1,\dots,Y_r\},
    \label{eq:def_principal_sector_simple}
\end{equation}
with \(\mathcal{W}_0^\star=\{0\}\). 
No other \(r\)-dimensional subspace yields a smaller value of \(I_1\).

For this canonical choice, the residual of each operator word is
\begin{equation}
    R_m^{(\mathcal{W}_r^\star)}=(\mathbb{I}-\mathbb{P}_{\mathcal{W}_r^\star})X_m=\sum_{\alpha>r}\sqrt{\lambda_\alpha}V_{m\alpha}^*Y_\alpha.\label{eq:principal_residual_simple}
\end{equation}
The residual Gram matrix is therefore
\begin{equation}
    Q^{(\mathcal{W}_r^\star)}_{mn}=\sum_{\alpha>r}\lambda_\alpha V_{m\alpha}V_{n\alpha}^*,
    \label{eq:principal_residual_gram_entries_simple}
\end{equation}
where we used the orthonormality of the modes \(Y_\alpha\). 
Equivalently,
\begin{equation}
    Q^{(\mathcal{W}_r^\star)}(\Omega)
    =
    V\operatorname{diag}
    (0,\dots,0,\lambda_{r+1},\dots,\lambda_N)V^\dagger .
    \label{eq:principal_residual_gram_simple}
\end{equation}
Thus, under canonical conditioning, the irreducible geometry is governed entirely by the tail spectrum \(\lambda_{r+1},\dots,\lambda_N\). 
Using the spectral expression in Eq.~\eqref{eq:Dq_elementary_symmetric}, we obtain
\begin{equation}
    I_q(\Omega,\mathcal{W}_r^\star)
    =
    \sum_{r+1\le i_1<\cdots<i_q\le N}
    \lambda_{i_1}\cdots\lambda_{i_q},
    \label{eq:principal_profile}
\end{equation}
for \(q=1,\dots,N-r\), with \(I_q(\Omega,\mathcal{W}_r^\star)=0\) for \(q>N-r\). 
A direct calculation after normalization further gives
\begin{equation}
    \bar{q}_r=\sum_{\alpha=r+1}^{N}\frac{\lambda_\alpha}{1+\lambda_\alpha},
    \quad
    (\Delta q_r)^2=\sum_{\alpha=r+1}^{N}\frac{\lambda_\alpha}{(1+\lambda_\alpha)^2}.
    \label{eq:canonical_qbar}
\end{equation}
The overall procedure is summarized schematically in Fig.~\ref{fig2}(a).

\subsection{Pauli-string family}
\label{sec:3:subsec:B}

We first consider a simple benchmark in which \(\Omega=\{X_m\}_{m=1}^{N}\) consists of hermitian Pauli strings. 
Since Pauli operators either commute or anticommute, distinct operator words can reduce to the same Pauli string up to an overall sign.
We therefore group the elements of \(\Omega\) by the Pauli string they represent, ignoring the overall sign, and denote the number of distinct groups by \(N_P\).
We denote a representative Pauli string in group \(\mu\) by \(P_\mu\) and the size of that group by \(n_\mu\).

For \(X_m\) in the \(\mu\)-th group, we write \(X_m=\eta_m P_\mu\) with \(\eta_m=\pm1\). 
If \(X_m\) belongs to the \(\mu\)-th group and \(X_n\) belongs to the \(\nu\)-th group, then
\begin{equation}
    \langle X_m,X_n\rangle=\eta_m^*\eta_n\delta_{\mu\nu}.
\end{equation}
Thus, operators in different groups are orthogonal, while operators in the same group have an overlap of \(\pm 1\).

After reordering the labels so that operators in the same group are adjacent, the Gram matrix takes the block form
\begin{equation}
    G(\Omega)=
    \begin{pmatrix}
        G_1 & 0 & \cdots & 0\\
        0 & G_2 & \cdots & 0\\
        \vdots & \vdots & \ddots & \vdots\\
        0 & 0 & \cdots & G_{N_P}
    \end{pmatrix}.
\end{equation}
The \(\mu\)-th block is an \(n_\mu\times n_\mu\) matrix with entries \((G_\mu)_{mn}=\eta_m^*\eta_n\). 
Since this block has rank one, it has a single nonzero eigenvalue \(n_\mu\), with all remaining eigenvalues equal to zero. 
Consequently, the nonzero eigenvalues of the full Gram matrix are \(n_1,\dots,n_{N_P}\).
The corresponding principal modes are the Pauli-string directions themselves, and canonical conditioning resolves these directions in order of decreasing multiplicity.
Substituting the corresponding residual spectrum into Eqs.~\eqref{eq:principal_profile} and \eqref{eq:canonical_qbar} then gives the canonical irreducible volume profile, which is therefore controlled solely by the multiplicities of the Pauli-string directions.

This benchmark illustrates two limiting cases. 
If \(N_P=1\), all operator words reduce to the same Pauli string up to sign. 
The full Gram matrix has rank one, and the correlator geometry is spanned by a single direction. 
Conversely, if every element of \(\Omega\) belongs to a distinct Pauli-string group, then \(N_P=N\), \(G(\Omega)=\mathbb{I}\), and all eigenvalues are equal to one. 
In this uniform limit, all directions contribute equally to the correlator geometry, leaving no distinguished low-dimensional structure to select.

\subsection{Haar-randomized family}
\label{sec:3:subsec:C}

We next consider a family of random operator words.
Let the operator words \(X_m\) be independently Haar-randomized, for example by setting \(X_m=U_m^\dagger X_m' U_m\) with independent Haar-random unitaries \(U_m\). 
We decompose each operator word into its identity and traceless components,
\begin{equation}
    X_m=c_m \mathbb{I} + B_m,
    \quad
    c_m=\frac{1}{d}\TrOp X_m,
    \quad
    \TrOp B_m=0,
\end{equation}
with normalization \( |c_m|^2+\|B_m\|^2=1\). 
The Gram matrix then takes the form
\begin{equation}
    G_{mn}=c_m^*c_n+\delta_{mn}\|B_m\|^2+\delta G_{m\neq n},
    \label{eq:haar_general_gram_typical}
\end{equation}
where the off-diagonal fluctuation \(\delta G_{m\neq n}\) has typical size \(O(d^{-1})\).

In the traceless case, where \(c_m=0\) for all \(m\), the Gram matrix is close to the identity, with off-diagonal corrections of order \(O(d^{-1})\). 
Accordingly, its spectrum is nearly flat:
\begin{equation}
    \lambda_\alpha=1+O(d^{-1}).
    \label{eq:haar_spectrum}
\end{equation}
Thus, traceless Haar-randomized operator words approach the uniform limit described above. 
Note that when the identity components \(c_m\) are nonzero, the rank-one term \(c_m^*c_n\) can produce an \(O(1)\) contribution to the Gram matrix, and the spectrum need not remain flat.

In the uniform limit, the irreducible volume profile becomes binomial:
\begin{equation}
    I_q(\Omega,\mathcal{W}_r^\star)=\binom{N-r}{q}+O(d^{-1}),
\end{equation}
and
\begin{equation}
    \pi_q(\Omega,\mathcal{W}_r^\star)
    =\frac{1}{2^{N-r}}\binom{N-r}{q}+O(d^{-1}),
\end{equation}
for \(q=0,\dots,N-r\). 
The mean and variance are
\begin{equation}
    \bar{q}_r=\frac{N-r}{2}+O(d^{-1}),
    \quad
    (\Delta q_r)^2=\frac{N-r}{4}+O(d^{-1}).
    \label{eq:Haar_reference}
\end{equation} 
We use this profile as the uniform-limit benchmark for a maximally spread irreducible geometry.

\subsection{Free, interacting integrable, and chaotic dynamics}
\label{sec:3:subsec:D}

\begin{figure*}[t!]
    \centering
    \includegraphics[width=\hsize]{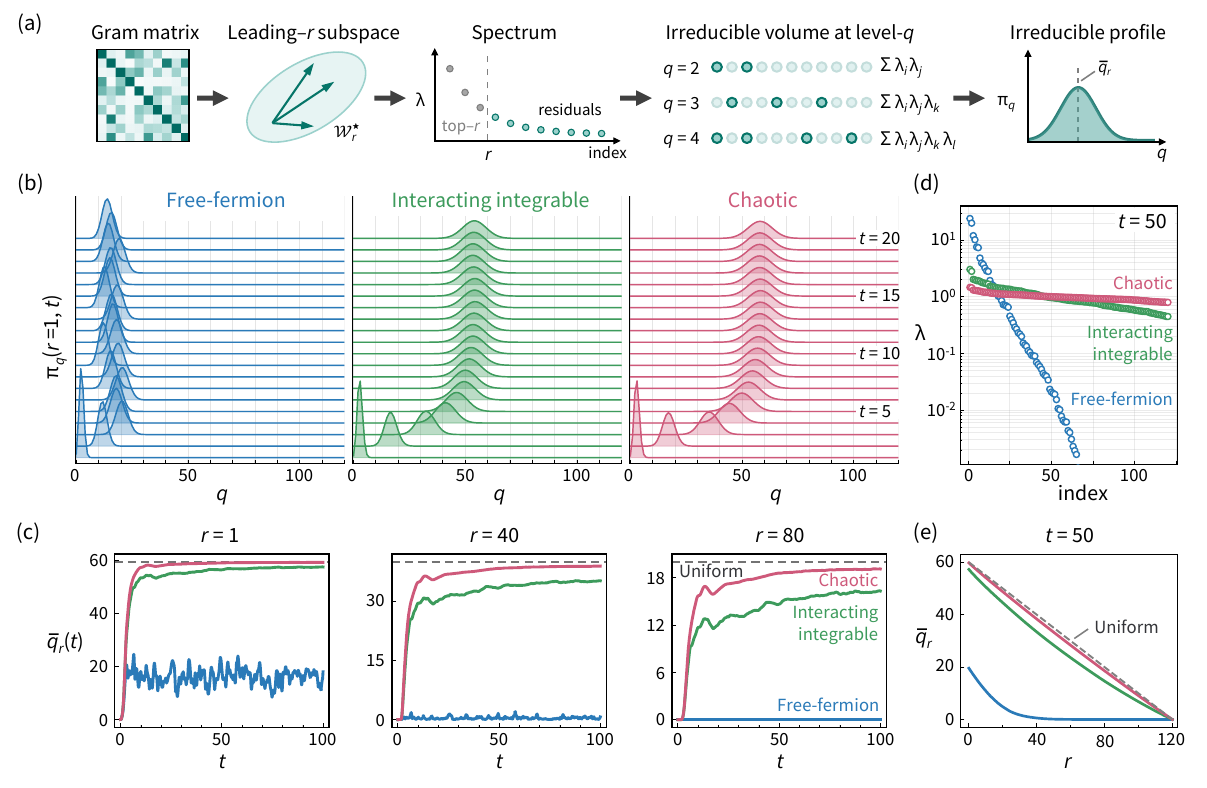}
    \caption{
        (a) Schematic procedure for constructing the irreducible volume profile under canonical conditioning. 
        (b) Time dependence of the irreducible volume profiles \(\pi_q(r,t) = \pi_q (\Omega(t), \mathcal{W}_r^\star(t))\) for the operator word set in Eq.~\eqref{eq:probe}, shown at fixed conditioning rank \(r=1\). 
        The three panels show free-fermion, interacting integrable, and chaotic dynamics at system size \(L=9\). 
        Each ridge corresponds to a different base time \(t\). 
        (c) Time evolution of the mean irreducible level \(\bar{q}_r(t)\) for representative conditioning ranks \(r\). 
        (d) Eigenvalue spectra of the Gram matrix \(G(\Omega)\) at the representative late time \(t=50\).
        (e) Conditioning-rank dependence of the mean irreducible level \(\bar{q}_r(t)\) at \(t=50\). 
        The dashed lines in (c) and (e) denote the corresponding uniform-limit references.
    }\label{fig2}
\end{figure*}

We now use canonical conditioning to ask whether different classes of many-body dynamics generate distinct higher-order correlator geometries. 
We consider spin-\(1/2\) XYZ chains with open boundary conditions
\begin{equation}
    H=\sum_{i=1}^{L-1}\left(J_x \sigma_i^x \sigma_{i+1}^x + J_y \sigma_i^y \sigma_{i+1}^y + J_z \sigma_i^z \sigma_{i+1}^z\right)+\sum_{i=1}^{L}\left(h_x \sigma_i^x + h_z \sigma_i^z\right),
    \label{eq:XYZ_Hamiltonian}
\end{equation}
at system size \(L=9\).
We compare three representative parameter sets for \((J_x,J_y,J_z,h_x,h_z)\):
\begin{equation}
    \begin{aligned}
        \text{free-fermion:} \quad & (1.0,0.7,0.0,0.0,0.6) \\
        \text{interacting integrable:} \quad
        &(1.0,0.7,0.4,0.0,0.0) \\
        \text{chaotic:} \quad &(1.0,0.7,0.4,0.3,0.6). 
    \end{aligned}
    \label{eq:XYZ_parameter}
\end{equation}
The first model is mapped to free fermions by the Jordan-Wigner transformation. 
The second model is interacting but remains integrable, and is solvable by the Bethe ansatz~\cite{Baxter:PRL:1971,Baxter:AnnPhys:1973}.
The third model is a generic nonintegrable model exhibiting quantum chaotic behavior~\cite{Shiraishi:EPL:2019}.

For each model, we construct the operator word family
\begin{equation}
    \Omega(t)=\operatorname{Perm}\{\sigma^x_1(0), \sigma^z_3(0.1t), \sigma^z_5(0.3t), \sigma^z_7(0.7t), \sigma^y_9(t)\}.
    \label{eq:probe}
\end{equation}
Here, \(\operatorname{Perm}\) denotes all ordered products obtained by permuting the five listed operators. 
Thus \(|\Omega|=N=5!\), and each \(X_m(t)\) is an operator word of length five.
For example,
\begin{equation}
    \begin{aligned}
        X_1(t)=\sigma^x_1(0)\sigma^z_3(0.1t)\sigma^z_5&(0.3t)\sigma^z_7(0.7t)\sigma^y_9(t),  \\
        X_2(t)=\sigma^z_3(0.1t)\sigma^x_1(0)\sigma^z_5&(0.3t)\sigma^z_7(0.7t)\sigma^y_9(t),\\
        &\vdots \\
        X_N(t)=\sigma^y_9(t)\sigma^z_7(0.7t)\sigma^z_5&(0.3t)\sigma^z_3(0.1t)\sigma^x_1(0).
    \end{aligned}\label{eq:perms}
\end{equation}
Pairing two such operator words produces a Gram matrix element \(\langle X_m(t),X_n(t)\rangle\), which is generally a ten-point correlator. 
For this representative choice, the five operator sites are equally spaced across the chain, the Pauli axes include several spin components rather than a single common axis, and the time arguments are chosen as nonuniform fractions of the base time \(t\) to avoid unintended interference effects.
Since each base operator is unitary, the corresponding operator words are already normalized in the Hilbert-Schmidt norm.
We also note that, although each individual Heisenberg Pauli operator remains traceless under time evolution, their product \(X_m(t)\) is not necessarily traceless.
Appendix~\ref{Appendix:Paulis} presents robustness checks with respect to the choice of operator words and system size.

Figure~\ref{fig2}(b) shows the time evolution of the irreducible volume profile under canonical conditioning at \(r=1\). 
At early times, the profile is sharply concentrated near small \(q\). 
The initially local operators have not yet spread significantly, and exchanging two operators within an operator word changes the word only slightly.
Therefore, different orderings of the five operators remain highly redundant.
Even a small canonical sector then captures much of the correlator family, leaving only a low-dimensional irreducible geometry. 
As time increases, the operators spread, and their noncommutativity makes different operator words increasingly distinct. 
The residual components then span more independent directions, producing a broader irreducible volume profile and a larger mean irreducible level \(\bar{q}_r(t)\).

Figure~\ref{fig2}(c) makes the distinction among the three dynamical regimes more quantitative by showing \(\bar{q}_r(t)\) for \(r=1,40,\) and \(80\). 
In the free-fermion model, \(\bar{q}_r(t)\) remains far below the uniform benchmark throughout the evolution. 
As the conditioning rank increases, \(\mathcal{W}_r^\star\) resolves nearly all of the \(120\) operator words, leaving \(\bar{q}_r(t)\) close to zero. 
This indicates that the correlator family is effectively organized by only a few collective directions. 
In contrast, the interacting integrable and chaotic models develop substantially broader irreducible geometries, with \(\bar{q}_r(t)\) rising rapidly toward the uniform value \((N-r)/2\).
Notably, the value for the interacting integrable model remains systematically below that for the chaotic model. 
While interactions can generate a high-dimensional irreducible sector, integrability still constrains this spreading through conserved structures that are absent in the chaotic model.

Figures~\ref{fig2}(d) and \ref{fig2}(e) provide further evidence for this geometric distinction at a representative late time \(t=50\), showing the Gram matrix eigenvalue spectrum and the conditioning rank dependence of the mean irreducible level, respectively.
In Fig.~\ref{fig2}(d), the eigenvalue spectrum of \(G(\Omega)\) decays rapidly in the free-fermion model, is flatter but still visibly sloped in the interacting integrable model, and becomes nearly flat in the chaotic model.
Since each eigenvalue measures the weight carried by a principal mode, this gradual flattening across dynamical regimes provides complementary spectral evidence that appreciable weight is distributed over increasingly many directions.
Figure~\ref{fig2}(e) further shows that this distinction persists across conditioning ranks \(r\). 
In the free-fermion model, \(\bar{q}_r\) becomes nearly zero already around \(r\simeq 40\), whereas in the two interacting models it decreases much more slowly with \(r\). 
The chaotic profile remains close to the uniform baseline over a wide range of ranks, while the interacting integrable profile stays systematically lower, consistent with integrability constraints.
Taken together, these results demonstrate that different dynamical regimes organize the same correlator family into qualitatively distinct geometries.

Previous studies have often characterized dynamical regimes through the typical behavior of individual correlators, including the growth, decay, and long-time behavior of OTOCs~\cite{Maldacena:JHEP:2016,Hosur:JHEP:2016,Lin:PRB:2018}. 
The emergence of such typical patterns may seem to suggest that the correlator family is largely redundant, as if it were governed by only a few representative forms. 
Our results reveal that this inference is incomplete: at the geometric level, typicality appears as the approach of the correlator geometry to the uniform limit, where the residuals spread over many independent directions and no few typical directions are singled out.
The comparison with free-fermion and interacting integrable dynamics further shows that this geometric typicality is a dynamical feature that emerges most strongly in the chaotic regime, rather than a generic consequence of considering a large correlator family.
Consequently, this perspective establishes correlator geometry as a higher-level description of many-body dynamics: dynamical features are encoded not only in individual correlator behavior, but also in the geometric organization of the full correlator family.

\section{Targeted conditioning geometry}
\label{sec:4}

In Sec.~\ref{sec:3}, we introduced canonical conditioning, in which the conditioning sector is determined intrinsically from the operator word family. 
Here, we turn to targeted conditioning, in which the conditioning subspace \(\mathcal{W}\) is fixed a priori based on physical considerations.
In Sec.~\ref{sec:4:subsec:A}, we use spatially targeted conditioning to identify the component of the correlator geometry that depends on a chosen spatial region in MBL dynamics.
In Sec.~\ref{sec:4:subsec:B}, we introduce measurement-targeted conditioning, where the conditioning sector is defined by the operator components accessible to a fixed measurement basis.
In Sec.~\ref{sec:4:subsec:C}, we extend the construction to state-dependent Gram matrices and use energy-resolved conditioning to uncover how the choice of state is reflected in the resulting geometry.
These examples show that targeted conditioning turns irreducible volume profiles into physically interpretable diagnostics of higher-order correlator structure.

\subsection{Spatially targeted conditioning}
\label{sec:4:subsec:A}

Conventional studies often characterize operator spreading through the spatial growth of OTOC profiles, typically by tracking the propagation and broadening of their wavefronts~\cite{Chen:AnnPhys:2017,Nahum:PRX:2018,Keyserlingk:PRX:2018,Khemani:PRX:2018}.
Here, we extend this perspective from individual correlators to the geometry of an entire higher-order correlator family.
To this end, we use spatially targeted conditioning: a region \(A\) is fixed a priori, and the conditioning procedure isolates the component of the correlator geometry that depends on \(A\).
The resulting irreducible volume profile reveals the spatial structure of the correlator geometry by characterizing its \(A\)-dependent component.

Let \(A\) be the target region and \(A^c\) its complement. 
We define the spatial conditioning sector by treating the operator algebra outside \(A\) as resolved:
\begin{equation}
    \mathcal{W}_{A^c}=\mathcal{B}(\mathcal{H}_{A^c})\otimes \mathbb{I}_A.
\end{equation}
The Hilbert-Schmidt projection onto this sector is 
\begin{equation}
    \mathbb{P}_{\mathcal{W}_{A^c}}X_m=\frac{1}{d_A}\TrOp_A(X_m)\otimes \mathbb{I}_A,
\end{equation}
where \(d_A=\dim\mathcal{H}_A\), and \(\TrOp_A\) denotes the partial trace over \(A\). 
Thus, \(\mathbb{P}_{\mathcal{W}_{A^c}}X_m\) is the component of \(X_m\) accessible from the complement \(A^c\), whereas the residual \(R_m^{(\mathcal{W}_{A^c})}\) is the irreducible component that requires nontrivial access to the target region \(A\). 
We diagnose this spatial structure of the correlator family through its irreducible geometry: \(I_1(\Omega,\mathcal{W}_{A^c})\) measures the total volume of the \(A\)-dependent irreducible component, while \(\bar{q}\) characterizes the geometric organization of this contribution.

\begin{figure}[t]
    \centering
    \includegraphics[width=\hsize]{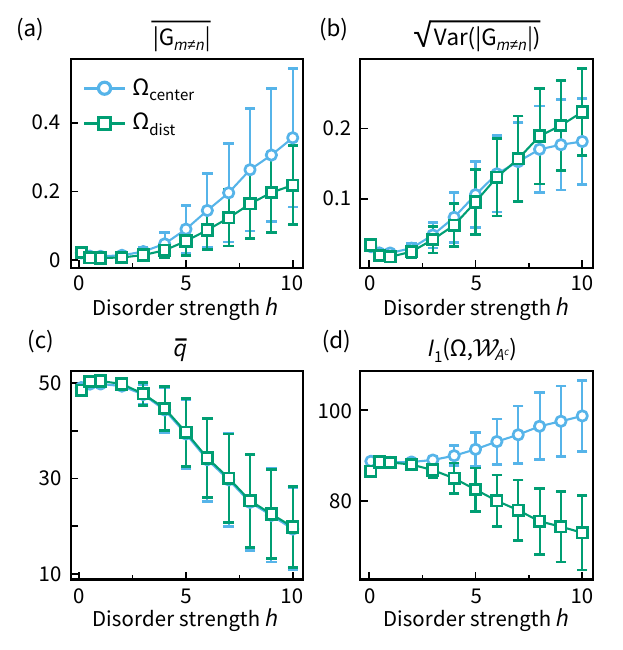}
    \caption{
        (a) Disorder strength dependence of the mean absolute value of the off-diagonal correlators \(\overline{|G_{m\neq n}|}\), and (b) the corresponding standard deviation \(\sqrt{\operatorname{Var}(|G_{m\neq n}|)}\) for the MBL Hamiltonian in Eq.~\eqref{eq:MBL_Hamiltonian}. 
        The mean and standard deviation are first computed over all off-diagonal entries \(m\neq n\) of the Gram matrix, and these quantities are then averaged over disorder realizations.
        (c) Mean irreducible level \(\bar{q}\) and (d) first irreducible volume \(I_1(\Omega, \mathcal{W}_{A^c})\) as functions of disorder strength \(h\).  
        The conditioning sector is chosen as \(\mathcal{W}_{A^c}\), where \(A=\{5\}\) is the target region.
        In all panels, the blue and green curves show results for the operator families \(\Omega_{\mathrm{center}}\) and \(\Omega_{\mathrm{dist}}\), respectively. 
        Data are averaged over \(500\) disorder realizations, and error bars indicate the standard deviation across disorder realizations.
    }\label{fig3}
\end{figure}

We apply this conditioning to the random-field Heisenberg spin-\(1/2\) chain,
\begin{equation}
    H=\sum_{i=1}^{L-1}J\left(\sigma_i^x\sigma_{i+1}^x +\sigma_i^y \sigma_{i+1}^y+\sigma_i^z\sigma_{i+1}^z\right)+\sum_{i=1}^{L} h_{z,i} \sigma_i^z,
    \label{eq:MBL_Hamiltonian}
\end{equation}
at system size \(L=9\).
The fields \(h_{z,i}\) are drawn independently from the uniform distribution \(h_{z,i}\in[-h,h]\). 
For sufficiently large disorder strength \(h\), this model realizes the MBL regime~\cite{Znidaric:PRB:2008,Pal:PRB:2010,Bardarson:PRL:2012,Huse:PRB:2014,Luitz:PRB:2015,Rahul:AnnRev:2015,Chen:AnnPhys:2017,Abanin:RevModPhys:2019}.

We choose the central site \(A=\{5\}\) as the target region and use its complement \(\mathcal{W}_{A^c}\) as the conditioning subspace. 
We then compare two operator word families. 
The first is
\begin{equation}
    \Omega_{\mathrm{center}}=\operatorname{Perm}\{\sigma^z_5(0), \sigma^x_5(0.3t), \sigma^z_5(t), \sigma^x_1(0.1t), \sigma^y_9(0.7t)\},
\end{equation}
where several base operators are initially supported on the target site \(A\). 
This family probes how operators originating from \(A\) contribute to the \(A\)-dependent correlator geometry.
The second is
\begin{equation}
    \Omega_{\mathrm{dist}}=\operatorname{Perm}\{\sigma^x_1(0), \sigma^z_2(0.1t), \sigma^y_3(0.3t), \sigma^x_7(0.7t), \sigma^y_9(t)\},
\end{equation}
where the base operators are distributed across the chain and initially avoid the target site. 
This family probes whether \(A\)-dependent geometry can be generated dynamically even when no operator is initially supported at \(A\).
In both cases, we evaluate the correlator family at the late time \(t=50\).

We begin by examining raw statistics of the individual correlators in Figs.~\ref{fig3}(a) and \ref{fig3}(b).
For each disorder realization, we compute the mean and standard deviation of \(|G_{m\neq n}|\) over all off-diagonal pairs, and then average these quantities over \(500\) disorder samples.
At the level of these raw statistics, \(\Omega_{\mathrm{center}}\) and \(\Omega_{\mathrm{dist}}\) exhibit the same overall trend: both the mean magnitude and the standard deviation increase with disorder strength. 
The growth of the mean magnitude would suggest stronger retention of initial information, whereas the enhanced standard deviation might be read as a greater diversity of correlator behavior.
The similarity would further suggest no qualitative difference between the two operator-word families.

A different picture emerges once the same correlators are analyzed at the level of correlator geometry.
In Fig.~\ref{fig3}(c), we present the disorder-strength dependence of the mean irreducible level \(\bar{q}\).
For both \(\Omega_{\mathrm{center}}\) and \(\Omega_{\mathrm{dist}}\), \(\bar{q}\) decreases gradually with \(h\), revealing that the \(A\)-dependent residual geometry becomes more redundant and effectively lower-dimensional as localization strengthens.
Therefore, the growth of the entrywise variance in Fig.~\ref{fig3}(b) should not be taken as evidence for more independent \(A\)-sensitive information across the correlator family. 
Rather, it reflects only a broader distribution of individual correlator values.

The geometric analysis further reveals a clear distinction between the two families in the \(A\)-dependent component.
As shown in Fig.~\ref{fig3}(d), the irreducible volume \(I_1(\Omega,\mathcal{W}_{A^c})\) decreases with \(h\) for the distributed family \(\Omega_{\mathrm{dist}}\). 
This indicates that operators initially supported away from \(A\) become increasingly unable to generate an \(A\)-dependent contribution as localization strengthens.
Conversely, \(I_1(\Omega,\mathcal{W}_{A^c})\) increases with \(h\) for the centered family \(\Omega_{\mathrm{center}}\). 
Since several base operators in this family are initially supported at \(A\), stronger localization prevents their contribution from spreading away from \(A\), leaving a larger \(A\)-dependent irreducible volume.

Spatially targeted conditioning therefore uncovers many-body structure that remains hidden in raw correlator summaries. 
This identifies the lower-dimensional organization and support-selective confinement of the correlator geometry as a geometric signature of MBL.

\subsection{Measurement-targeted conditioning}
\label{sec:4:subsec:B}

We next apply targeted conditioning in an experimentally motivated setting, in which the chosen measurement basis determines which operator components are directly accessible.
For example, projective measurements in the computational basis access the diagonal operator algebra generated by \(\mathbb{I}_i\) and \(\sigma_i^z\), whereas components that are off-diagonal in this basis, such as those involving \(\sigma_i^x\) or \(\sigma_i^y\), remain inaccessible without additional basis rotations or coherent measurement protocols. 
The measurement basis therefore defines a natural conditioning subspace, allowing us to separate the correlator geometry accessible with the chosen measurement basis from the geometry hidden by that basis.

We formalize this idea by introducing measurement-accessible subspaces.
For a Pauli string \(P=P_1\cdots P_L\), we define its off-diagonal Pauli weight relative to the computational basis by
\begin{equation}
    w_{\perp}(P)=
    \# \{i:\ P_i=\sigma_i^x\ {\mathrm{or}}\ P_i=\sigma_i^y \}.
    \label{eq:def_transverse_weight}
\end{equation}
The measurement sector with off-diagonal weight cutoff \(k\) is
\begin{equation}
    \mathcal{W}^{\mathrm{meas}}_{\le k}=\operatorname{span}
    \left\{P \mid w_{\perp}(P)\le k\right\}.
    \label{eq:def_measurement_sector}
\end{equation}
The sector \(\mathcal{W}^{\mathrm{meas}}_{\le 0}\) consists of Pauli strings built only from \(\mathbb{I}_i\) and \(\sigma_i^z\), and hence represents the purely diagonal component in the computational basis.
Increasing \(k\) progressively includes operator components with more off-diagonal support. 
At \(k=L\), all Pauli strings are included, so the sector coincides with the full operator space.

We use \(\mathcal{W}^{\mathrm{meas}}_{\le k}\) as the conditioning subspace for the operator word family \(\Omega=\{X_m\}_{m=1}^{N}\).
The first irreducible volume \(I_1(\Omega,\mathcal{W}^{\mathrm{meas}}_{\le k})\) then measures the part of the correlator geometry that remains inaccessible to measurements restricted to off-diagonal depth at most \(k\).
The irreducible volume profile \(\pi_q(\Omega,\mathcal{W}^{\mathrm{meas}}_{\le k})\) and its mean level \(\bar{q}\) further characterize the geometric organization of this measurement-inaccessible component.
The resulting \(k\)-dependence allows us to track the progressive resolution of the correlator geometry as higher off-diagonal components are included.

We demonstrate the correlator geometry using the mixed-field Ising Hamiltonian
\begin{equation}
    H=J\sum_{i=1}^{L-1} \sigma^z_i \sigma^z_{i+1}+g\sum_{i=1}^{L} \sigma^x_i+h\sum_{i=1}^{L} \sigma^z_i,
    \label{eq:mixed_ising_measurement}
\end{equation} 
at system size \(L=9\). 
The three terms play complementary roles with respect to the computational basis.
The longitudinal field proportional to \(h\) is diagonal and one-body, and is therefore directly resolved by the computational-basis measurement.
The interaction proportional to \(J\) is also diagonal in this basis, but it spreads information across sites.
The transverse field proportional to \(g\) generates off-diagonal components and is therefore the direct source of measurement-inaccessible components.
For the operator word family, we choose
\begin{equation}
    \Omega(t)=\operatorname{Perm}
    \left\{\sigma^z_1(0), \sigma^z_3(0.1t),\sigma^z_5(0.3t), \sigma^z_7(0.7t),\sigma^z_9(t)\right\}.
    \label{eq:measurement_probe_allZ}
\end{equation}
At \(t=0\), all base operators are contained in \(\mathcal{W}^{\mathrm{meas}}_{\le 0}\). 
Any measurement-inaccessible component is therefore generated dynamically by the transverse field.

\begin{figure}[t]
    \centering
    \includegraphics[width=\hsize]{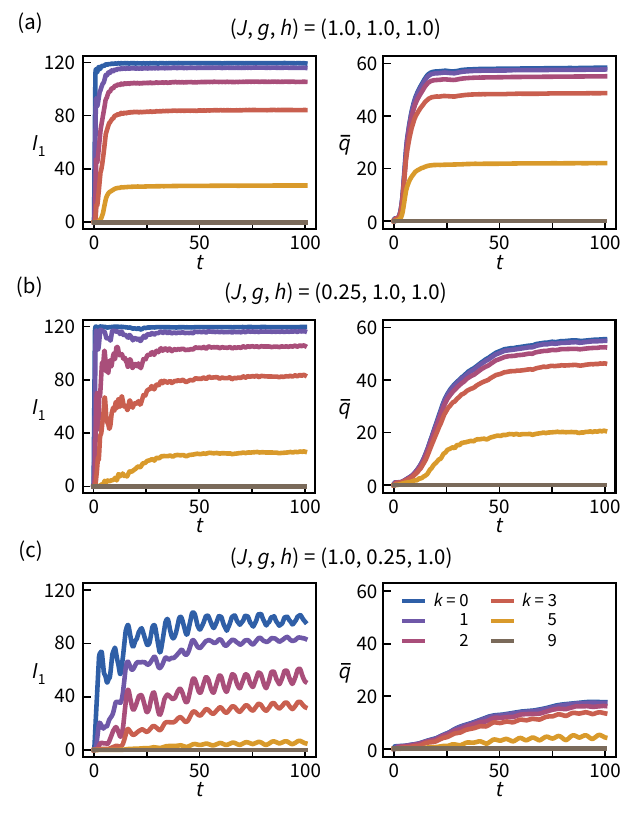}
    \caption{
        Time evolution of the first irreducible volume \(I_1(\Omega,\mathcal{W}^{\mathrm{meas}}_{\le k})\) and the mean irreducible level \(\bar{q}\), obtained by conditioning with respect to the measurement-accessible subspace \(\mathcal{W}^{\mathrm{meas}}_{\le k}\) for several off-diagonal depths \(k\).
        The mixed-field Hamiltonian in Eq.~\eqref{eq:mixed_ising_measurement} is employed with parameters (a) \((J, g, h) = (1.0, 1.0, 1.0)\), (b) \((0.25, 1.0, 1.0)\), and (c) \((1.0, 0.25, 1.0)\). 
        The operator word family \(\Omega\) is constructed from the all \(\sigma^z\) probe in Eq.~\eqref{eq:measurement_probe_allZ}. 
    }\label{fig4}
\end{figure}

In Fig.~\ref{fig4}, we show the time evolution of the first irreducible volume \(I_1(\Omega,\mathcal{W}^{\mathrm{meas}}_{\le k})\) and the mean irreducible level \(\bar{q}\) for three parameter regimes.
We first consider the generic interacting case \((J,g,h)=(1.0,1.0,1.0)\), shown in Fig.~\ref{fig4}(a).
For small \(k\), \(I_1\) rapidly approaches its maximal value \(N=120\), while \(\bar{q}\) also grows to large values.
This indicates that low-depth measurement restrictions leave an inaccessible component that is not only large in total weight but also has a high-dimensional residual geometry.
As \(k\) is increased, conditioning on the enlarged measurement-accessible sector resolves this residual structure more completely, with both \(I_1\) and \(\bar{q}\) decreasing and eventually vanishing at \(k=L\).

We now isolate the roles of the interaction and the transverse field by reducing each term in turn.
Figure~\ref{fig4}(b) shows the weak interaction case \((J,g,h)=(0.25,1.0,1.0)\). 
The first irreducible volume again grows rapidly, while the growth of \(\bar{q}\) is noticeably delayed compared with the generic interacting case.
Nevertheless, \(\bar{q}\) eventually reaches large values, suggesting that weaker interactions mainly delay the formation of the high-dimensional residual geometry rather than qualitatively altering its late-time structure.
Figure~\ref{fig4}(c) shows the case with a weak transverse field, \((J,g,h)=(1.0,0.25,1.0)\). 
Although the Hamiltonian is dominated by terms that are diagonal in the computational basis, a substantial measurement-inaccessible component still develops, as indicated by the large value of \(I_1\).
However, the mean level \(\bar{q}\) remains much smaller than in the previous two regimes, even at late times.
The pronounced oscillations in \(I_1\) further suggest that the inaccessible component is repeatedly exchanged with the accessible sector, rather than rapidly spreading over many residual directions.

Consequently, the comparison in Fig.~\ref{fig4} shows that generating measurement-inaccessible components and organizing them into many independent residual directions are distinct processes, controlled in different ways by the Hamiltonian terms.
Measurement inaccessibility is therefore not merely a loss of signal at the level of individual correlators.
Rather, inaccessibility itself has a geometry: correlators become inaccessible collectively, forming a mutually organized structure.

\subsection{State-dependent targeted conditioning}
\label{sec:4:subsec:C}

Thus far, we have formulated correlator geometry using the Hilbert-Schmidt inner product in Eq.~\eqref{eq:HS_inner}. 
This corresponds to an infinite-temperature average over operator space, which treats all states in the Hilbert space on equal footing.
In some settings, however, correlators are defined relative to a specified reference state, such as an energy eigenstate~\cite{Hashimoto:JHEP:2017}, a thermal state~\cite{Maldacena:JHEP:2016}, or an experimentally prepared initial state~\cite{Google:Nature:2025}.
This motivates the state-dependent correlator geometry studied here, in which overlaps between operator words are evaluated with respect to the chosen reference state.
The conditioning subspace can then be selected either by canonical optimization or on the basis of physically relevant state properties.

The construction begins with a pure reference state \(|\psi\rangle\).
For each operator word \(X_m\), we define the normalized state generated from \(|\psi\rangle\) by
\begin{equation}
    |X_m(t,\psi)\rangle = \frac{X_m(t)|\psi\rangle}{\sqrt{\langle\psi|X_m^\dagger(t)X_m(t)|\psi\rangle}} .
\label{eq:state_descendant_energy_window}
\end{equation}
The corresponding state-dependent Gram matrix is
\begin{equation}
    G_{mn}(\Omega(t),\psi) = \langle X_m(t,\psi)|X_n(t,\psi)\rangle.
    \label{eq:state_resolved_gram_energy_window}
\end{equation}
Equivalently, before normalization, this construction replaces the Hilbert-Schmidt inner product by the state-dependent inner product
\begin{equation}
    \langle X,Y\rangle_\psi=\langle\psi|X^\dagger Y|\psi\rangle=\TrOp(\rho X^\dagger Y),
\end{equation}
with \(\rho=|\psi\rangle\langle\psi|\). 
The Hilbert-Schmidt geometry is recovered by taking \(\rho=\mathbb{I}/d\).

The conditioning procedure is analogous to that in the operator case, with the conditioning sector now taken to be a subspace of the Hilbert space.
Let \(\mathcal{W}\subset\mathcal{H}\) be a conditioning subspace and let \(P_{\mathcal{W}}: \mathcal{H}\rightarrow\mathcal{H}\) denote the orthogonal projector onto it. 
Each operator-generated state is then decomposed as
\begin{equation}
    |X_m(t,\psi)\rangle=P_{\mathcal{W}}|X_m(t,\psi)\rangle+|R_m^{(\mathcal{W})}(t,\psi)\rangle .
    \label{eq:state_sector_decomposition}
\end{equation}
The residual Gram matrix is given by
\begin{equation}
    Q_{mn}^{(\mathcal{W})}(\Omega(t),\psi)=
    \langle R_m^{(\mathcal{W})}(t,\psi) | R_n^{(\mathcal{W})}(t,\psi) \rangle .
\label{eq:state_residual_gram}
\end{equation}
The irreducible volumes \(I_q(\Omega,\mathcal{W})\) and the normalized profile \(\pi_q(\Omega,\mathcal{W})\) are defined from this residual Gram matrix in exactly the same way as in Sec.~\ref{sec:2:subsec:B}.

\begin{figure}[t]
    \centering
    \includegraphics[width=\linewidth]{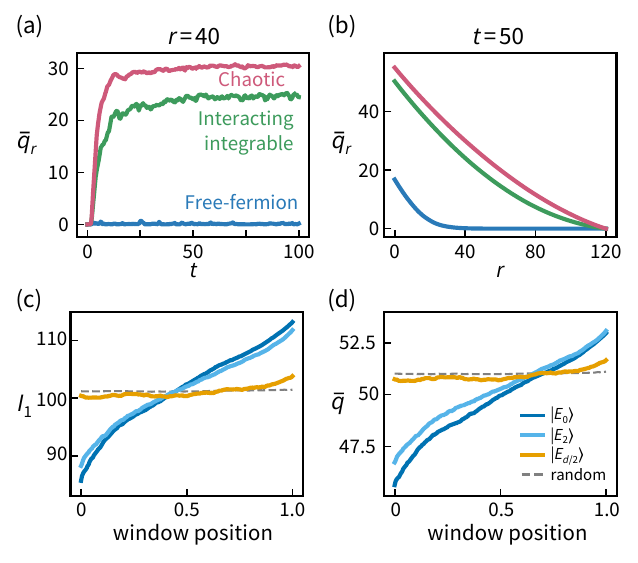}
    \caption{
        (a) Time evolution of the mean irreducible level \(\bar{q}_r(t)\) for the state-dependent geometry with the ground state \(|E_0\rangle\) as the reference state, under canonical conditioning at fixed rank \(r=40\).
        The Hamiltonian and parameters are defined in Eqs.~\eqref{eq:XYZ_Hamiltonian} and \eqref{eq:XYZ_parameter}, and the operator word family \(\Omega\) is defined in Eq.~\eqref{eq:probe}.
        (b) Dependence of \(\bar{q}_r\) on the canonical conditioning rank \(r\) at the representative late time \(t=50\). 
        The blue, green, and red curves in (a) and (b) correspond to the free-fermion, interacting integrable, and chaotic models, respectively.
        (c) First irreducible volume \(I_1(\Omega,\mathcal{W}^{(j,r)})\) and (d) mean irreducible level \(\bar{q}\) for energy-resolved conditioning in the chaotic XYZ model at \(t=50\) and \(r=80\).
        The horizontal axis represents the normalized spectral window position \(j/(d-r)\), where \(\mathcal{W}^{(j,r)}\) is the rank-\(r\) target conditioning subspace spanned by consecutive energy eigenstates beginning with \(|E_j\rangle\).
        Different curves in (c) and (d) correspond to \(|\psi\rangle=|E_0\rangle\), \(|E_2\rangle\), \(|E_{d/2}\rangle\), and a random-state average.
    }
    \label{fig5}
\end{figure}

We apply this state-dependent construction to the XYZ spin chains in Eq.~\eqref{eq:XYZ_Hamiltonian}, using the same operator word family \(\Omega\) as in Eq.~\eqref{eq:probe} at system size \(L=9\). 
The parameter sets for the free-fermion, interacting integrable, and chaotic models are defined in Eq.~\eqref{eq:XYZ_parameter}. 
For each model, we diagonalize the Hamiltonian and denote its eigenstates by \(|E_a\rangle\), ordered as \(E_0\le E_1\le\cdots\).

We first construct the state-dependent Gram matrix using the ground state \(|E_0\rangle\) as the reference state, and perform canonical conditioning as in Sec.~\ref{sec:3:subsec:A}. 
Figure~\ref{fig5}(a) shows the time evolution of the mean irreducible level \(\bar{q}_r(t)\) for the three models at fixed rank \(r=40\), while Fig.~\ref{fig5}(b) shows its dependence on the canonical conditioning rank \(r\) at \(t=50\).
These panels are the state-dependent counterparts of the infinite-temperature results in Figs.~\ref{fig2}(c) and \ref{fig2}(e), respectively.
In the free-fermion model, \(\bar{q}_r\) drops almost to zero by \(r\simeq 40\) in Fig.~\ref{fig5}(b), and correspondingly remains close to zero in Fig.~\ref{fig5}(a), suggesting that the irreducible geometry is effectively low-dimensional.
In contrast, the interacting integrable and chaotic models develop broader irreducible profiles with rapid growth of \(\bar{q}_r\) in Fig.~\ref{fig5}(a). 
The chaotic curve lies consistently above the interacting integrable curve in Fig.~\ref{fig5}(b), indicating that integrability constrains the state-dependent correlator geometry.
Consequently, distinct dynamical regimes generate distinct correlator geometries in the state-dependent setting as well, consistent with the results for the Hilbert-Schmidt geometry discussed in Sec.~\ref{sec:3:subsec:D}.

The close correspondence suggests that the dependence on the reference state can be largely obscured under canonical conditioning.
This is because the canonical conditioning subspace is optimized separately for each state-dependent Gram matrix, so that much of the change in the inner product can be absorbed into the optimized subspace. 
Consistent with this interpretation, we also confirm that the irreducible profiles remain qualitatively similar when the reference state is chosen to be another energy eigenstate or a random state.
Note that this does not imply that the correlator geometry is always insensitive to the choice of reference state. 
Such dependence can become more apparent under targeted conditioning, where the conditioning sector is fixed independently of the Gram matrix.

Among several possible targeted choices, we choose conditioning sectors from the spectral decomposition of the Hamiltonian.
We define a rank-\(r\) conditioning window starting at energy index \(j\) as 
\begin{equation}
    \mathcal{W}^{(j,r)}=\operatorname{span}\{|E_j\rangle,|E_{j+1}\rangle,\ldots,|E_{j+r-1}\rangle\}.
\end{equation}
The normalized window position is given by \(j/(d-r)\) with \(d=\dim\mathcal{H}\). 
The lowest-energy rank-\(r\) sector corresponds to \(j/(d-r)=0\), while the highest-energy rank-\(r\) sector corresponds to \(j/(d-r)=1\).
Conditioning on \(\mathcal{W}^{(j,r)}\) therefore probes the organization of the state-dependent geometry relative to a fixed energy sector.

Figures~\ref{fig5}(c) and \ref{fig5}(d) show the first irreducible volume \(I_1\) and the mean irreducible level \(\bar{q}\), respectively, as functions of the energy window position. 
For the chaotic XYZ model, we compare state-dependent geometries generated from \(|\psi\rangle=|E_0\rangle\), \(|E_2\rangle\), and the middle spectrum eigenstate \(|E_{d/2}\rangle\) at \(t=50\) with the conditioning rank \(r=80\).
We also include an average over \(100\) randomly sampled states as a baseline. 
This random-state reference shows little dependence on the window position, consistent with the absence of a preferred energy sector.

For the correlator geometries generated from \(|E_0\rangle\) and \(|E_2\rangle\), both \(I_1\) and \(\bar{q}\) are minimized when the conditioning sector is built from the low-lying part of the spectrum.
They increase as the same fixed-rank window is shifted toward the middle or upper part of the spectrum, and even exceed the random-state baseline once the window is sufficiently high in energy. 
This identifies the low-energy window as the natural target sector for low-energy reference states, resolving their irreducible structure more efficiently than spectrally distant windows.
In contrast, for the middle-spectrum eigenstate \(|E_{d/2}\rangle\), sliding the window across the spectrum reveals no comparable energy sector preference: \(I_1\) and \(\bar{q}\) remain close to the random-state baseline over most window positions, apart from modest deviations near the high-energy part of the spectrum.
In this way, energy-resolved conditioning demonstrates that the reference state leaves a geometric imprint on the correlator family, most clearly through the energy-sector selectivity of low-lying eigenstates.

It is important to note that there are several inequivalent ways of introducing state dependence into correlators.
For instance, in the context of thermal OTOCs~\cite{Maldacena:JHEP:2016}, one usually considers OTOCs of the form \(\TrOp (\rho^{1/4} W(t)\rho^{1/4} V \rho^{1/4} W(t)\rho^{1/4} V)\), which are generally distinct from \(\TrOp (\rho W(t) V W(t) V)\).
Our construction is closer in spirit to the latter form, in which state dependence enters through the modified inner product that defines the geometry of operator words.
The former type can also be accommodated within our framework by shifting state dependence from the inner product to the choice of operator word family itself.
For example, when an operator word \(X_m\) is constructed as a product of \(M\) operators, one may absorb the density matrix by dressing each operator as \(O_j\mapsto \rho^{1/(4M)}O_j\rho^{1/(4M)}\), while keeping the Hilbert-Schmidt inner product.
The resulting correlator geometries should generally be inequivalent, and we leave their systematic comparison for future work.

\section{Comparing correlator geometries}
\label{sec:5}

In the preceding sections, we treated a correlator family as a geometry in operator space and showed that this geometry can encode physical structure that is invisible at the level of individual correlator values. 
So far, however, we have analyzed one correlator family at a time. 
We now ask how correlator geometries can be compared as geometric objects, beyond comparisons of their irreducible volume profiles. 
To this end, we develop two distinct comparison schemes.
In Sec.~\ref{sec:5:subsec:A}, we compare correlator geometries generated by the same dynamics at different times.
This leads to Krylov conditioning, which compares time-separated correlator geometries through Hamiltonian-generated directions.
In Sec.~\ref{sec:5:subsec:B}, we consider correlator geometries generated by different dynamics.
For this more general setting, we introduce cross conditioning, in which the canonical sector of one correlator family is used as the conditioning sector for another. 

\subsection{Krylov conditioning}
\label{sec:5:subsec:A}

We first compare correlator geometries generated by the same Hamiltonian at different times. For a fixed choice of operator word family, the time-dependent family \(\Omega(t)\) defines a trajectory of geometries in operator space.
Comparing \(\Omega(t_1)\) and \(\Omega(t_2)\) therefore probes the dynamical reshaping of the correlator geometry.

We begin with the simplest way to compare two geometries, namely by measuring the overlap between their canonical conditioning subspaces.
We consider the chaotic XYZ Hamiltonian in Eqs.~\eqref{eq:XYZ_Hamiltonian} and \eqref{eq:XYZ_parameter}, together with the operator word family \(\Omega\) in Eq.~\eqref{eq:probe}. 
The leading rank-\(r\) canonical subspace at time \(t\) is
\begin{equation}
    \mathcal{W}_r^\star(t)=\operatorname{span}\{Y_1(t),\ldots,Y_r(t)\},
\end{equation}
where \(Y_\alpha(t)\) are the principal modes of the Gram matrix \(G(\Omega(t))\). 
We then quantify the overlap between the canonical subspaces at two times \(t_1\) and \(t_2\) as
\begin{equation}
    C(t_1,t_2)=\frac{1}{r}
    \sum_{\alpha,\beta=1}^{r}
    \left|\langle Y_\alpha(t_1),Y_\beta(t_2)\rangle\right|^2 .
\end{equation}
By construction, \(0\le C(t_1,t_2)\le1\), with \(C=1\) when the two rank-\(r\) subspaces coincide.

\begin{figure}[t]
    \centering
    \includegraphics[width=\hsize]{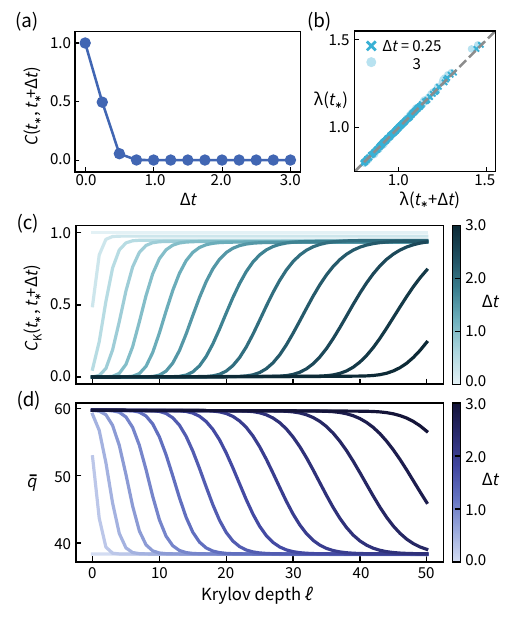}
    \caption{
    (a) Direct overlap \(C(t_*, t_*+\Delta t)\) between the leading \(r=40\) canonical subspaces at the reference time \(t_*=50\) and at the later time \(t_*+\Delta t\). 
    We use the chaotic XYZ chain defined in Eqs.~\eqref{eq:XYZ_Hamiltonian} and \eqref{eq:XYZ_parameter}, together with the operator word set \(\Omega\) defined in Eq.~\eqref{eq:probe}. 
    (b) Scatter plot comparing the canonical Gram spectra \(\lambda(t_*)\) and \(\lambda(t_*+\Delta t)\) for \(\Delta t = 0.25\) and \(3\). 
    (c) Krylov-resolved overlap \(C_\mathrm{K}(t_*,t_*+\Delta t;r,\ell)\) as a function of the Krylov depth \(\ell\) for \(r=40\) and \(t_*=50\). 
    (d) Mean irreducible level \(\bar{q}\) after conditioning \(\Omega(t_*+\Delta t)\) by the Krylov sector \(\mathcal{W}_{\mathrm{K}}^{(r,\ell)}(t_*)\). 
    Colors in (c) and (d) indicate \(\Delta t\), sampled in increments of \(0.25\).
    }\label{fig6}
\end{figure}

In Fig.~\ref{fig6}(a), we present \(C(t_*,t_*+\Delta t)\) as a function of \(\Delta t\), with the reference time fixed at \(t_*=50\).
Notably, the direct subspace overlap decays rapidly with \(\Delta t\) and becomes nearly zero already at small time separations.
In Fig.~\ref{fig6}(b), we also compare the Gram spectra at \(t_*\) and \(t_*+\Delta t\) for two representative separations \(\Delta t=0.25\) and \(3\); the former still retains a finite value of \(C(t_*,t_*+\Delta t)\), whereas the latter has almost zero overlap.
For both separations, the spectra remain almost unchanged relative to the reference spectrum and therefore give nearly identical irreducible volume profiles.
These results suggest that the correlator geometry is substantially reoriented within operator space without appreciably changing the internal organization of the operator words.
However, such a change does not by itself identify the dynamical relation between the two geometries, since a similar pattern could also arise from a global rotation of the entire geometry. 
Because the diagnostics above compare the two geometries only as static objects, they cannot distinguish a displacement generated by the Hamiltonian from a generic reorientation in operator space. 
This motivates a comparison that explicitly incorporates Hamiltonian evolution into the conditioning procedure.

We implement this idea by enlarging the reference canonical sector with Hamiltonian-generated Krylov directions.
For a seed operator \(O\), the standard Krylov sequence is generated by repeated action of the Liouvillian \(\mathcal{L}\):
\begin{equation}
    \{O,\mathcal{L}[O], \mathcal{L}^2[O],\dots\},
    \quad
    \mathcal{L}[O]= [H,O].
\end{equation}
In the present setting, however, the seeds are not single operators but products of operators with several time arguments.
A slight generalization is therefore required.

For simplicity, we consider an operator word family \(\Omega(t)\) generated by permuting \(M\) base operators whose time arguments are parameterized by a common variable \(t\): \(\Omega (t) = \operatorname{Perm}\{O_1(c_1 t),\dots,O_M(c_M t)\}\) with fixed coefficients \(\{c_i\}\). 
For a permutation \(\sigma\in\mathfrak{S}_M\), we write
\begin{equation}
    X_\sigma(t)=O_{\sigma(M)}(c_{\sigma(M)}t)\cdots O_{\sigma(1)}(c_{\sigma(1)}t).
    \label{eq:word_path_definition_Lnotation}
\end{equation}
Let \(\mathcal{L}^{[s]}\) denote the operation that inserts the Liouvillian into the \(s\)-th slot of the word,
\begin{align}
    \mathcal{L}^{[s]}
    \left[O_{\sigma(M)}\cdots O_{\sigma(s)} \cdots O_{\sigma(1)}\right]=O_{\sigma(M)}\cdots \mathcal{L}[O_{\sigma(s)}] \cdots O_{\sigma(1)}.
    \label{eq:slot_liouvillian_product}
\end{align}
We then define the Liouvillian associated with the ordering \(\sigma\) as
\begin{equation}
    \mathcal{L}_{\sigma}=\sum_{s=1}^{M}c_{\sigma(s)}\mathcal{L}^{[s]} .
    \label{eq:path_liouvillian_sigma}
\end{equation}
It generates the evolution of the whole operator word with respect to \(t\):
\begin{equation}
    X_{\sigma}(t+\Delta t)
    =\sum_{m=0}^{\infty}\frac{(i\Delta t)^m}{m!}\mathcal{L}_{\sigma}^m X_{\sigma}(t).
\end{equation}

Using the principal-mode expansion of the operator word in Eq.~\eqref{eq:def_principal_modes_simple}, we define the \(m\)-th Krylov evolution of \(Y_\alpha(t)\) by applying the corresponding ordering-dependent Liouvillian to each operator word component as
\begin{equation}
    \mathcal{K}^{m}Y_\alpha(t)=\frac{1}{\sqrt{\lambda_\alpha}}
    \sum_{\sigma}V_{\sigma\alpha}(t)\mathcal{L}_{\sigma}^{m}X_\sigma(t).
    \label{eq:krylov_descendant_principal_mode}
\end{equation}
For \(m=0\), this reduces to the original principal mode \(Y_\alpha(t)\).
We then define the rank-\(r\), depth-\(\ell\) Krylov conditioning sector as
\begin{equation}
    \mathcal{W}_{\mathrm{K}}^{(r,\ell)}(t)
    =\operatorname{span}
    \left\{
        \mathcal{K}^{m}Y_\alpha(t):
        \ 1\le \alpha \le r,
        \ 
        \ 0\le m \le \ell
    \right\}.
    \label{eq:alpha_krylov_sector_Lnotation}
\end{equation}
When \(c_1=\cdots=c_M\), all operators in each word are evaluated at a common Heisenberg time, and the construction reduces to the usual Krylov construction or to its multi-seed extension~\cite{Parker:PRX:2019,Craps:PRL:2025}.

We now quantify the overlap between the later canonical conditioning subspace \(\mathcal{W}_r^\star(t_2)\) and the Krylov sector generated from the earlier time \(\mathcal{W}_{\mathrm{K}}^{(r,\ell)}(t_1)\). 
Let \(\{Z_j^{(r,\ell)}(t_1)\}_{j=1}^{d^{(r,\ell)}}\) be an orthonormal basis of \(\mathcal{W}_{\mathrm{K}}^{(r,\ell)}(t_1)\), where \(d^{(r,\ell)} \) denotes its dimension. 
The Krylov-resolved overlap is given by
\begin{equation}
    C_{\mathrm{K}}(t_1,t_2;r,\ell) = \frac{1}{r} \sum_{\alpha=1}^{r} \sum_{j=1}^{d^{(r,\ell)}}
    \left|
        \left\langle
            Z_j^{(r,\ell)}(t_1),Y_\alpha(t_2)
        \right\rangle
    \right|^2,
\end{equation}
where \(0\leq C_{\mathrm{K}}(t_1,t_2;r,\ell)\leq1\). 
When \(C_{\mathrm{K}}(t_1,t_2;r,\ell) \simeq 0\), the leading \(r\) canonical conditioning subspace of the later geometry is almost invisible from the depth-\(\ell\) Krylov sector generated from the earlier geometry. 
In contrast, \(C_{\mathrm{K}}(t_1,t_2;r,\ell) \simeq 1\) suggests that this later canonical subspace is almost fully captured by the same Krylov-expanded sector.

Figure~\ref{fig6}(c) shows the Krylov-resolved overlap \(C_\mathrm{K}(t_*,t_*+\Delta t;r,\ell)\) for the chaotic XYZ model as a function of the Krylov depth \(\ell\) with \(r=40\) and \(t_*=50\).
At \(\ell=0\), the Krylov sector is just the reference canonical sector itself, and the overlap is correspondingly small. 
As \(\ell\) increases, the Krylov sector incorporates additional Hamiltonian-generated directions, and \(C_\mathrm{K}\) begins to grow.
The onset of this growth occurs at larger Krylov depths for larger time separations, whereas smaller \(\ell\) is sufficient for shorter \(\Delta t\).
The dynamical reshaping of the geometry is reflected in this systematic dependence on \(\Delta t\).
At sufficiently large \(\ell\), \(C_\mathrm{K}\) reaches a plateau.
This plateau remains close to, yet below, one, indicating that the Krylov-expanded sector generated by the leading \(r\) reference modes captures a large fraction, but not all, of the leading canonical subspace at later times.
The remaining part can be interpreted as leakage into directions outside the Krylov sector, arising from mixing with subleading canonical directions during the Hamiltonian evolution.

The same trend appears in the geometric diagnostic obtained by conditioning \(\Omega(t_*+\Delta t)\) with respect to the Krylov sector \(\mathcal{W}_{\mathrm{K}}^{(r,\ell)}(t_*)\).
In Fig.~\ref{fig6}(d), we present the mean irreducible level \(\bar{q}\) as a function of the Krylov depth \(\ell\) with \((r,t_*)=(40,50)\). 
For small \(\ell\), the mean irreducible level remains close to the high-dimensional value associated with an almost unconditioned target geometry, indicating that the Krylov sector has not yet captured the correlator geometry at \(t_*+\Delta t\). 
As \(\ell\) increases, \(\bar{q}\) drops toward the value at \(\Delta t = 0\), namely the canonical conditioning value obtained after removing the leading \(r\) modes.
The decrease of \(\bar{q}\) therefore indicates that the Krylov-expanded reference sector increasingly resolves the later correlator geometry, providing a geometric counterpart to the growth of \(C_{\mathrm{K}}\) in Fig.~\ref{fig6}(c).

Consequently, Krylov conditioning reveals a Hamiltonian-generated relation between correlator geometries at different times, which is not apparent from either the direct subspace overlap or the individual geometry profiles. 
This construction thereby extends Krylov analysis from single-operator spreading to the geometric reshaping of an entire higher-order correlator family under many-body dynamics.

\subsection{Cross conditioning}
\label{sec:5:subsec:B}

\begin{figure*}[t!]
    \centering
    \includegraphics[width=\hsize]{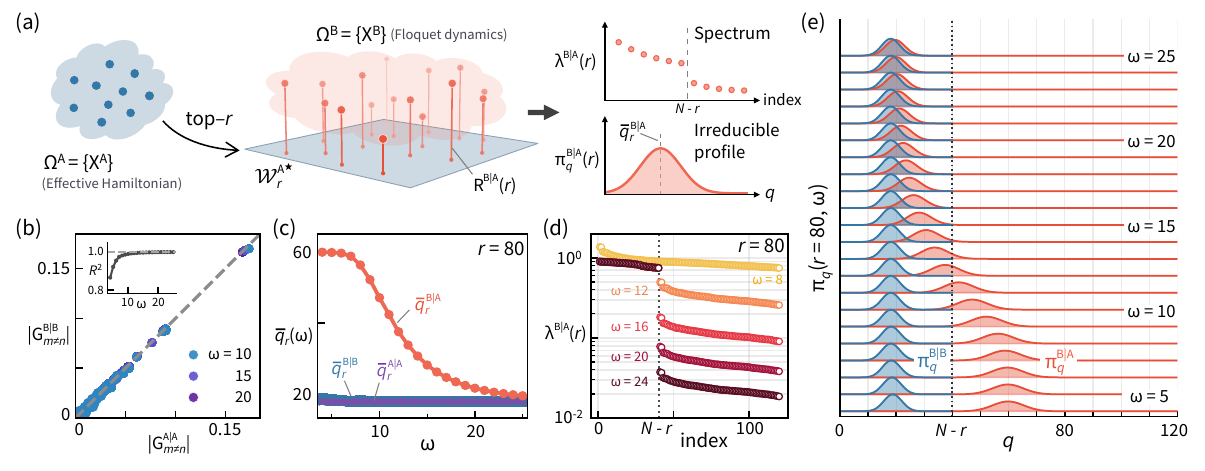}
    \caption{
        (a) Schematic of the cross conditioning procedure.
        (b) Scatter plot comparing the absolute values of the off-diagonal Gram matrix elements obtained from the effective-Hamiltonian dynamics \(|G^{A|A}_{m\neq n}|\) and from the Floquet dynamics \(|G^{B|B}_{m\neq n}|\).        
        Different colors correspond to different drive frequencies \(\omega=2\pi/T\). 
        The inset shows the squared Pearson correlation \(R^2\) between the two sets of off-diagonal correlators as a function of \(\omega\).
        (c) Frequency dependence of the mean irreducible levels \(\bar{q}_r^{A|A}\), \(\bar{q}_r^{B|B}\), and \(\bar{q}_r^{B|A}\) at conditioning rank \(r=80\).  
        (d) Eigenvalue spectra \(\lambda^{B|A}(r)\) of the residual Gram matrix \(Q^{B|A}(r)\) for several frequencies \(\omega\). 
        The conditioning rank is \(r=80\), and the dotted vertical line marks \(N-r\).
        (e) Irreducible volume profiles at \(r=80\) as a function of \(\omega\). 
        Blue and orange ridges show \(\pi_q^{B|B}\) and \(\pi_q^{B|A}\), respectively. 
    }\label{fig7}
\end{figure*}

We now introduce a more general comparison scheme, cross conditioning, which compares correlator geometries generated by different dynamics.
The central idea is to use the canonical conditioning sector of one correlator family as the reference sector for another.
This yields a directed diagnostic to quantify how much of the target geometry is explained by the dominant geometric directions of the reference family.

Let \(\Omega^A=\{X_m^A\}_{m=1}^{N_A}\) and \(\Omega^B=\{X_m^B\}_{m=1}^{N_B}\) denote two operator word families; we use \(A\) as the reference and \(B\) as the target. 
First, we construct the \(r\)-dimensional canonical conditioning subspace \(\mathcal{W}_r^{A\star}\) from \(\Omega^A\). 
The target operator words are then decomposed as  
\begin{equation}
    X_m^B=\mathbb{P}_{{\mathcal{W}_r^{A\star}}}X_m^B+R_m^{B|A}(r),
\end{equation}
where \(R_m^{B|A}(r)\perp \mathcal{W}_r^{A\star}\).
The notation \(B|A\) emphasizes that the comparison is directed.
The corresponding cross residual Gram matrix is
\begin{equation}
    Q_{mn}^{B|A}(r)=\langle R_m^{B|A}(r),R_n^{B|A}(r)\rangle,
    \label{eq:cross_residual_gram}
\end{equation}
which characterizes the geometric mismatch from \(A\) to \(B\).

Writing \(\lambda_1^{B|A}(r),\dots,\lambda_{N_B}^{B|A}(r)\) for the eigenvalues of \(Q^{B|A}(r)\),  we define the cross-conditioned irreducible volumes by
\begin{equation}
    I_q^{B|A}(r) = \sum_{1\le i_1<\cdots<i_q\le N_B}\lambda_{i_1}^{B|A}(r)\cdots\lambda_{i_q}^{B|A}(r),
    \quad
    q=1,\dots, N_B, \label{eq:cross_I}
\end{equation}
with \(I_0^{B|A}(r)=1\).
The normalized profile \(\pi_q^{B|A}(r)\) and its mean \(\bar{q}_r^{B|A}\) are defined accordingly; the cross conditioning procedure is schematically illustrated in Fig.~\ref{fig7}(a).
To distinguish these cross-conditioned profiles from the canonical ones, we write \(\pi_q^{A|A}(r)\) and \(\pi_q^{B|B}(r)\) for the usual canonical conditioning profiles of \(A\) and \(B\), respectively.
Note that for the canonical conditioning of \(\Omega^B\), the leading \(r\) principal directions of the target geometry are removed, so that \(\pi_q^{B|B}(r)=0\) for \(q>N_B-r\).
In contrast, under cross conditioning, the reference sector \(\mathcal{W}_r^{A\star}\) need not coincide with the target canonical sector \(\mathcal{W}_r^{B\star}\).
Consequently, the sums in Eq.~\eqref{eq:cross_I} run over the full set of \(N_B\) eigenvalues, and the resulting profile \(\pi_q^{B|A}(r)\) can have nonzero weight for all \(q\).
Nevertheless, if \(\mathcal{W}_r^{A\star}\) is close to \(\mathcal{W}_r^{B\star}\), the leading \(r\) principal directions of \(\Omega^B\) are still nearly removed by \(\mathcal{W}_r^{A\star}\), so the sorted spectrum \(\{\lambda^{B|A}\}\) can exhibit a sharp drop near the effective rank \(N_B-r\).

We apply this framework to a periodically driven spin-\(1/2\) chain.
Based on the XYZ model in Eq.~\eqref{eq:XYZ_Hamiltonian}, we define two Hamiltonians
\begin{align}
    H_{+}&=H_{\mathrm{XYZ}}(J_x,J_y,J_z,+h_x^{\mathrm{drv}},h_z),\\
    H_{-}&=H_{\mathrm{XYZ}}(J_x,J_y,J_z,-h_x^{\mathrm{drv}},h_z),
    \label{eq:Hpm_xyz}
\end{align}
with \((J_x,J_y,J_z,\pm h_x^{\mathrm{drv}},h_z)=(1.0,0.7,0.4,\pm 0.3, 0.6)\).
We use them to define the Floquet time evolution as 
\begin{equation}
    U_F=e^{-iH_{-}T/2}e^{-iH_{+}T/2}
    \label{eq:UF_xyz}
\end{equation}
for one driving period \(T\). 
In the high-frequency limit, the Floquet-Magnus expansion gives the effective Hamiltonian~\cite{Blanes:PhysRep:2009}
\begin{equation}
    H_{\mathrm{eff}}=\frac{H_{+}+H_{-}}{2}-\frac{iT}{8}[H_{-},H_{+}] + O(T^2),
    \label{eq:Heff_xyz}
\end{equation}
where \(T=2\pi/\omega\), and the \(O(T^2)\) corrections are omitted.

In what follows, we use the effective-Hamiltonian evolution as the reference dynamics \(A\), and the exact Floquet evolution as the target dynamics \(B\).
The corresponding Heisenberg evolutions are
\begin{equation}
    O_j^A(t_j)=e^{iH_{\mathrm{eff}}t_j}O_je^{-iH_{\mathrm{eff}}t_j},
    \quad
    O_j^B(t_j)=U_F^{-n_j}O_jU_F^{n_j},
    \label{eq:AB_evolved_probes_xyz}
\end{equation}
evaluated at stroboscopic times \(t_j=n_jT\).
We then construct the operator word families from all permutations of the same five Pauli operators,
\begin{equation}
    \Omega^{A,B}=\operatorname{Perm}
    \left\{
        \sigma^x_1(t_1),
        \sigma^z_3(t_2),
        \sigma^z_5(t_3),
        \sigma^z_7(t_4),
        \sigma^y_9(t_5)
    \right\}.
    \label{eq:cross_probe}
\end{equation}
Each time-evolved operator is evaluated according to Eq.~\eqref{eq:AB_evolved_probes_xyz}, and the two families have the same size \(N=5!\).
We fix the physical time to \(t_*=80\) to probe the Floquet prethermal regime~\cite{Mori:PRL:2016}, and set
\begin{equation}
    n_j=\left\lfloor \frac{\alpha_j t_*}{T}\right\rfloor ,
    \quad
    \{\alpha_j\}=\{0,0.1,0.3,0.7,1\}.
\end{equation}
This prescription ensures that each probe is evaluated at the same stroboscopic time \(t_j=n_jT\) for both the discrete Floquet evolution and the continuous effective-Hamiltonian evolution.

We first compare the raw correlators generated by the two dynamics.
Figure~\ref{fig7}(b) shows the off-diagonal Gram matrix elements \(|G_{m\neq n}^{A|A}|=|\langle X_m^A,X_n^A\rangle|\) and \(|G_{m\neq n}^{B|B}|=|\langle X_m^B,X_n^B\rangle|\) for several drive frequencies. 
At high frequency, the two correlator sets nearly coincide, as expected from the construction of \(H_{\mathrm{eff}}\) as an effective description.
Upon lowering \(\omega\), corrections beyond the truncated Floquet-Magnus Hamiltonian are expected to become more important.
Even so, the squared Pearson correlation \(R^2\) stays close to one over a broad frequency range and remains above \(0.8\) in all regimes considered here.
In Fig.~\ref{fig7}(c), we show a complementary comparison based on the self-conditioned mean irreducible levels \(\bar{q}_r^{A|A}\) and \(\bar{q}_r^{B|B}\) evaluated at \(r=80\). 
These two quantities track each other closely as functions of \(\omega\), and both remain near the uniform reference value \((N-r)/2=20\).
Therefore, over much of the frequency range, the effective-Hamiltonian and Floquet correlator geometries remain nearly indistinguishable both at the level of individual correlators and at the level of self-conditioned volume profiles.

Cross conditioning reveals a different picture.
In Fig.~\ref{fig7}(c), we present the mean level of the cross-conditioned profile \(\bar{q}_r^{B|A}\) at the same conditioning rank \(r=80\).
Although \(\bar{q}_r^{B|A}\) is close to \(\bar{q}_r^{A|A}\) and \(\bar{q}_r^{B|B}\) at high frequency, it rises rapidly as \(\omega\) is lowered; the deviation becomes pronounced already around \(\omega\simeq20\), where the raw correlators still have \(R^2\simeq1\).
The spectral origin of this separation is shown in Fig.~\ref{fig7}(d), where we plot the sorted eigenvalues \(\lambda_i^{B|A}(r)\) of the residual Gram matrix.
For large \(\omega\), the spectrum exhibits a pronounced drop near \(i=N-r\), indicating that the reference sector \(\mathcal{W}_r^{A\star}\) removes approximately the same leading directions of \(\Omega^B\) as \(\mathcal{W}_r^{B\star}\) does.
Lowering \(\omega\) progressively softens this spectral drop, and by \(\omega=8\) the spectrum becomes nearly continuous, reflecting a gradual loss of alignment between the reference and target geometries.
Figure~\ref{fig7}(e) shows the corresponding volume profiles \(\pi_q^{B|A}\) and \(\pi_q^{B|B}\).
As \(\omega\) decreases, \(\pi_q^{B|A}\) evolves from a narrow low-\(q\) profile close to \(\pi_q^{B|B}\) into a broader profile shifted toward larger \(q\).
The shift reveals a breakdown of the effective description that becomes visible only through the relational geometry between the correlator families, even when the raw correlators and self-conditioned profiles remain nearly indistinguishable.

We are therefore led to distinguish three layers of structure in higher-order correlators.
The first consists of individual correlator values.
This layer tests direct numerical agreement, but discards the mutual organization of the correlators.
The second is the intrinsic geometry of a correlator family.
It treats the correlators as a structured family, but still analyzes each geometry in its own frame.
The third is relational: geometries are compared relative to a shared space.
Since the distinction between the effective and Floquet dynamics appears only at this relational level, correlator geometry should be viewed not only as an intrinsic representation of a single correlator family, but also as a relational object across different families.
Cross conditioning gives this relational viewpoint a concrete geometric form, completing the picture of correlator geometry as a higher-level description of many-body dynamics.

\section{Conclusions}
\label{sec:6}

In this work, we developed a geometric framework for analyzing higher-order correlators as structured families in operator space, rather than as individual numerical quantities.
The framework begins by viewing each correlator as an inner product between operator words, so that a full correlator family is encoded in an operator Gram matrix.
This perspective reveals structural features that are obscured at the level of raw correlator values, including redundancy, dependence, and genuinely high-dimensional many-body information.
We organize these relations by choosing a conditioning subspace \(\mathcal{W}\), which specifies the resolved sector.
This choice decomposes the correlator family into a reducible component explained by that sector and an irreducible component encoded in the residual geometry.
We then analyze the residual component not only through its norm, which measures the amount of irreducible information, but also through its geometry, which reveals how that information is structured.

We demonstrated this framework through several complementary forms of conditioning.
First, we introduced canonical conditioning, which provides an intrinsic and optimal way to explain a correlator family.
When applied to quantum spin dynamics, it revealed that different dynamical regimes imprint qualitatively distinct geometries on correlator families.
Second, we defined targeted conditioning, which uses physically motivated subspaces to give the resulting geometries specific interpretations.
Spatial conditioning distinguishes information retained in or propagated into a selected region.
Measurement conditioning reveals measurement inaccessibility as a structured geometry rather than mere signal loss.
Energy-resolved conditioning diagnoses spectral selectivity in state-dependent correlator geometries.
Finally, we extended the framework from individual correlator geometries to relations among geometries.
Krylov conditioning probes the displacement between correlator geometries at different times.
Cross conditioning instead compares geometries generated by different dynamics.
Together, these conditioning schemes establish correlator geometry as a higher-level framework for organizing, interpreting, and comparing many-body dynamics beyond individual correlator values.

Several avenues for future work follow naturally from this study.
One immediate direction is to connect correlator geometry more directly to dynamical phenomena that are usually diagnosed through individual observables, such as the growth of operator support~\cite{Lieb:1972}, front broadening~\cite{Keyserlingk:PRX:2018,Nahum:PRX:2018}, operator-entanglement growth~\cite{Zanardi:PRA:2001,Prosen:PRA:2007,Pizorn:PRB:2009}, and spreading in Krylov space~\cite{Parker:PRX:2019}.
Such a connection would turn these single-observable diagnostics into geometric statements about entire correlator families.
Another direction is to incorporate experimental limitations, such as noise, imperfect measurements, and finite-sample estimation, into the choice of conditioning sector.
This would recast limitations that are currently represented by scalar error diagnostics as geometric constraints on experimentally unresolved structure.
Finally, a more foundational route forward is to diagnose the limits of classical tractability through correlator geometry.
Recent measurements of higher-order OTOCs on programmable quantum processors suggest that some components of experimentally accessible correlator families may be difficult to reproduce classically~\cite{Google:Nature:2025}. 
This motivates decomposing a correlator family into a classically tractable sector and a residual sector that may encode genuinely quantum structure.
Such a decomposition could provide a geometric characterization of quantum advantage, although precisely defining the classically tractable sector remains a challenging problem.
Connections with quantum resource theories may offer a useful route toward this formulation~\cite{Chitambar:RevModPhys:2019}.
We leave these extensions to future work.

\begin{acknowledgments}
    We are grateful to Yukitoshi Motome for his comments on the manuscript.
    We thank Keisuke Fujii, Kazuaki Takasan, Ken Mochizuki, Keiya Sakabe, and Tomohiro Soejima for fruitful discussions.
    We also thank Frank Pollmann for collaborations on related topics.
    This work was supported by JST BOOST (No.~JPMJBS2418). 
\end{acknowledgments}

\textbf{Data availability}. 
Data analysis and simulation codes are available on Zenodo upon reasonable request~\cite{zenodo}.

\appendix

\section{Irreducible constraints on higher-order OTOCs}
\label{Appendix:OTOCs}

In this Appendix, we show how the Gram matrix construction applies to higher-order OTOCs~\cite{Google:Nature:2025, Fujii:arXiv:2025}.
Let \(A(t)\) and \(B\) be unitary operators, and define the higher-order OTOC moments by
\begin{equation}
    C^{(2m)}(t)=\frac{1}{d}\TrOp\left[(A(t)B)^{2m}\right].
    \label{eq:rep_otoc_moments}
\end{equation}
Since \(A(t)B\) is unitary, these moments satisfy
\begin{equation}
    C^{(-2m)}(t)=(C^{(2m)}(t))^* .
\end{equation}
For a finite moment depth \(M\), we introduce the operator word family
\begin{equation}
    X_m(t)=(A(t)B)^{2m},
    \quad
    m=0,1,\dots,M .
    \label{eq:rep_operator_words}
\end{equation}
The corresponding Gram matrix is
\begin{equation}
    G_{mn}(t) = \frac{1}{d}\TrOp\left[(A(t)B)^{2(n-m)}\right]
    =
    C^{(2(n-m))}(t),\label{eq:otoc_toeplitz_gram}
\end{equation}
or equivalently,
\begin{equation}
    G=
    \begin{pmatrix}
        1        & C^{(2)}      & C^{(4)}      & \cdots & C^{(2M)} \\
        (C^{(2)})^* & 1           & C^{(2)}     & \cdots & C^{(2(M-1))} \\
        (C^{(4)})^* & (C^{(2)})^*    & 1           & \cdots & C^{(2(M-2))} \\
        \vdots   & \vdots      & \vdots      & \ddots & \vdots \\
        (C^{(2M)})^* & (C^{(2(M-1))})^*& (C^{(2(M-2))})^*& \cdots & 1
    \end{pmatrix}.
    \label{eq:repeated_toeplitz_gram_explicit}
\end{equation}
Thus, the higher-order OTOC moments form a Toeplitz Gram matrix, whose entries depend only on the difference \(n-m\).

We now ask how the moment \(C^{(2M)}\) is constrained by the lower moments \(C^{(2)},\dots,C^{(2(M-1))}\).
For this purpose, we choose the lower-moment sector
\begin{equation}
    \mathcal{W}(t)=\operatorname{span}\left\{ X_1(t),X_2(t),\dots,X_{M-1}(t)\right\}
    \label{eq:otoc_lower_moment_sector}
\end{equation}
as the conditioning subspace.
The words \(X_1,\dots,X_{M-1}\) lie inside \(\mathcal{W}\), and hence their residuals vanish.
The only residuals that can be nonzero are therefore those of \(X_0\) and \(X_M\). 
The nontrivial part of the residual Gram matrix is the \(2\times2\) block 
\begin{equation}
    Q^{(\mathcal{W})} = 
    \begin{pmatrix} 
    \|R_0^{(\mathcal{W})}\|^2 & \langle R_0^{(\mathcal{W})},R_{M}^{(\mathcal{W})}\rangle \\ \langle R_{M}^{(\mathcal{W})},R_0^{(\mathcal{W})}\rangle & \|R_{M}^{(\mathcal{W})}\|^2 \end{pmatrix}. 
\end{equation} 
Its positivity gives
\begin{equation}
    |\langle R_{0}^{(\mathcal{W})},R_{M}^{(\mathcal{W})} \rangle|^2 \leq \|R_{0}^{(\mathcal{W})}\|^2\|R_{M}^{(\mathcal{W})}\|^2 .
    \label{eq:otoc_inequality}
\end{equation}
The residual overlap can further be written as
\begin{equation}
    \langle R_{0}^{(\mathcal{W})},R_{M}^{(\mathcal{W})} \rangle = C^{(2M)}-
    \left\langle
        \mathbb{P}_{\mathcal{W}}X_0,
        \mathbb{P}_{\mathcal{W}}X_{M}
    \right\rangle,
\end{equation}
which represents the part of the higher-order OTOC \(C^{(2M)}\) that is irreducible relative to the lower sector \(\mathcal{W}\).

To express this constraint explicitly in terms of lower moments, we define
\begin{equation}
    \begin{gathered}
        K_{mn}(t)=\langle X_m(t),X_n(t)\rangle,\\
        f_m(t) = \langle X_m(t),X_0(t)\rangle,
        \quad
        g_m(t) = \langle X_m(t),X_{M}(t)\rangle,
        \label{eq:otoc_Kk}
    \end{gathered}
\end{equation}
for \(m,n=1,\dots, M-1\). 
All entries of \(K\), \(\mathbf{f}\), and \(\mathbf{g}\) are therefore determined by the moments \(C^{(2)},\dots,C^{(2(M-1))}\) and their complex conjugates.
Using the standard projection formula, with \(K^+\) denoting the Moore-Penrose pseudoinverse, we obtain
\begin{equation}
    \left\langle
        \mathbb{P}_{\mathcal{W}}X_0,
        \mathbb{P}_{\mathcal{W}}X_{M}
    \right\rangle
    =\mathbf{f}^\dagger K^+ \mathbf{g},
    \label{eq:otoc_reducible_prediction_explicit}
\end{equation}
and
\begin{equation}
    \|R_{0}^{(\mathcal{W})}\|^2 = 1-\mathbf{f}^\dagger K^+ \mathbf{f},
    \quad
    \|R_{M}^{(\mathcal{W})}\|^2 = 1-\mathbf{g}^\dagger K^+ \mathbf{g}.
    \label{eq:otoc_residual_norm_compact}
\end{equation}
Together with Eq.~\eqref{eq:otoc_inequality}, these expressions show that the highest moment \(C^{(2M)}\) is constrained to lie within a disk in the complex plane whose center and radius are determined by the lower OTOC moments \(C^{(2)},\dots,C^{(2(M-1))}\).

\section{Robustness to operator words and system size}
\label{Appendix:Paulis}

In Sec.~\ref{sec:3:subsec:D}, we analyzed the correlator geometries with canonical conditioning for a representative operator word family \(\Omega\). 
Here, we demonstrate that our conclusions are robust to changes in both the choice of operator words and the system size.

We consider the XYZ Hamiltonian in Eq.~\eqref{eq:XYZ_Hamiltonian}, using the free-fermion, interacting integrable, and chaotic parameter sets defined in Eq.~\eqref{eq:XYZ_parameter}. 
For each system size \(L\), we independently sample \(500\) operator word families. 
Each family is generated from \(M=5\) base operators of the form \(\sigma_{i_a}^{\mu_a}(\alpha_a t)\), with \(a=1,\dots,5\). 
Here, the Pauli component \(\mu_a \in\{x,y,z\}\) is sampled randomly, while the site \(i_a \in \{1,\dots,L\}\) and the time fraction \(\alpha_a \in\{0,0.1,0.3,0.7,1.0\}\) are chosen without repetition among the five operators.  
The operator word family is then formed by taking all permutations of these base operators,
\begin{equation}
    \Omega(t)=\operatorname{Perm}
    \{\sigma_{i_1}^{\mu_1}(\alpha_1 t),\dots,\sigma_{i_5}^{\mu_5}(\alpha_5 t)\},
\end{equation}
so that \(N=M!\). 
For each sampled family, we compute the mean irreducible level \(\bar{q}_r\) under canonical conditioning at \(t=50\).
The resulting values of \(\bar{q}_r\) are summarized by their sample mean and standard deviation over this ensemble.

\begin{figure}[b]
    \centering
    \includegraphics[width=\hsize]{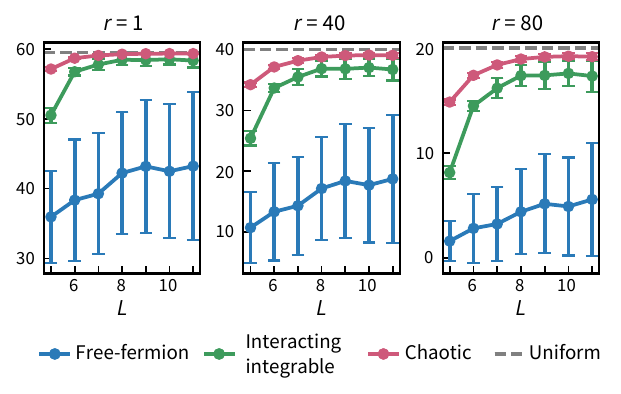}
    \caption{
        Mean irreducible level \(\bar{q}_r\) under canonical conditioning as a function of system size \(L\). 
        For each \(L\), the plotted values and error bars denote the sample mean and standard deviation of \(\bar{q}_r\) over \(500\) randomly sampled operator word families. 
        Results are shown for conditioning ranks \(r=1,40,\) and \(80\), evaluated at \(t=50\) for the free-fermion, interacting integrable, and chaotic parameter sets of the XYZ Hamiltonian defined in Eqs.~\eqref{eq:XYZ_Hamiltonian} and \eqref{eq:XYZ_parameter}.
    }\label{figA1}
\end{figure}

In Fig.~\ref{figA1}, we show the system size dependence of \(\bar{q}_r\) for conditioning ranks \(r=1,40,\) and \(80\). 
Notably, across the sampled operator word families and system sizes, the qualitative behavior observed in Fig.~\ref{fig2} remains unchanged. 
The chaotic dynamics produces a broadly spread irreducible geometry, with \(\bar{q}_r\) close to the uniform-limit reference. 
The interacting integrable dynamics also generates a large mean irreducible level, but it remains systematically below the chaotic value, consistent with constraints imposed by integrability. 
The free-fermion dynamics yields much smaller \(\bar{q}_r\), indicating that its correlator geometry is organized by a more restricted set of effective directions.

The fluctuations over choices of \(\Omega\) provide an additional diagnostic. 
In the chaotic model, the standard deviation is small, indicating that the correlator geometry approaches a typical form that is largely insensitive to microscopic details of the chosen operator words. 
The free-fermion model shows the opposite tendency: the standard deviation remains large, revealing a stronger dependence on the choice of operator words.
This behavior is consistent with the underlying quadratic structure, in which operator growth is constrained by fermionic dynamics rather than governed by generic many-body scrambling in operator space.
The interacting integrable model exhibits intermediate behavior, with fluctuations larger than in the chaotic case but substantially smaller than in the free-fermion case.

These results demonstrate that the conclusions drawn from the canonical conditioning analysis in Sec.~\ref{sec:3:subsec:D} are robust to changes in both the operator word family and the system size.
We note, however, that such robustness need not hold for targeted conditioning, since the conditioning sector is chosen to probe a specific physical feature that may itself depend on the chosen operator words.

\bibliography{bibtex}

\end{document}